\documentclass[11pt,a4paper,twocolumn]{article}

\usepackage[margin=1in]{geometry}
\usepackage{amsmath,amssymb}
\usepackage{graphicx}
\usepackage{booktabs}
\usepackage{array}
\usepackage{authblk}
\usepackage{microtype}
\microtypesetup{expansion=false}
\usepackage{makecell}
\usepackage{longtable}
\usepackage{etoolbox}
\makeatletter
\patchcmd{\LT@output}{\copy\LT@foot\vss}{%
  \copy\LT@foot\vskip 0pt plus \maxdimen minus \normalbaselineskip}{}{%
  \typeout{Warning: longtable infinite-glue patch failed}}
\patchcmd{\LT@output}{\copy\LT@foot\vss}{%
  \copy\LT@foot\vskip 0pt plus \maxdimen minus \normalbaselineskip}{}{%
  \typeout{Warning: longtable infinite-glue patch failed}}
\makeatother
\usepackage{url}
\usepackage{xurl}
\usepackage{float}
\usepackage{orcidlink}
\usepackage{hyperref}
\usepackage{cleveref}
\hypersetup{
  colorlinks=true,
  linkcolor=blue,
  citecolor=blue,
  urlcolor=blue
}

\newcommand{\GeV}{\text{GeV}}
\newcommand{\MET}{\ensuremath{E_{\mathrm{T}}^{\mathrm{miss}}}}

\newcommand{\invfb}{\ensuremath{\mathrm{fb}^{-1}}}
\newcolumntype{L}[1]{>{\raggedright\arraybackslash}p{#1}}

\title{%
  Hadronic Mono-\texorpdfstring{$Z$}{Z} Dark Matter Sensitivity with\\
  Flow Matching on CMS Open Data
}
\author[1,$\dagger$]{Hitesh Rasineni\,\orcidlink{0009-0003-4958-0915}\thanks{Corresponding author: \url{hitesh.23bce8825@vitapstudent.ac.in}}}
\author[2,$\dagger$]{Chebrolu Bhavishya\,\orcidlink{0009-0000-9770-0975}}

\affil[1]{VIT-AP University, Amaravati, 522241, India \\ \centerline{\small *hitesh.23bce8825@vitapstudent.ac.in}}
\affil[2]{Mohan Babu University, Tirupati, 517102, India \\ \centerline{\small 23102A041122@mbu.asia}}

\date{}

\begin{document}

\maketitle

\begingroup
\renewcommand{\thefootnote}{\fnsymbol{footnote}}
\footnotetext[2]{$\dagger$ These authors contributed equally and share first authorship.}
\endgroup
\begin{abstract}
We present a projected sensitivity study for hadronic mono-$Z$ dark-matter
production using CMS Run~2015D HTMHT open data corresponding to
2.256382381~\invfb, from which 1{,}439{,}523 events satisfy the hadronic
mono-$Z$ selection. Backgrounds are modelled with a conditional flow-matching
continuous normalizing flow trained on the selected HTMHT events and
evaluated on a held-out validation split reweighted to the full selected
population. To mitigate artifacts from missing-object features and avoid
in-sample scoring bias we apply sentinel imputation for undefined angular
features, persist the train/validation split indices, and enforce a minimum
reported background yield of 20 events when selecting the working point. A
signal-side offline trigger proxy is applied to the simulated signal before
scoring. Under this procedure the baseline analysis yields expected
significances of 2.89$\sigma$, 7.62$\sigma$, and 7.41$\sigma$ for three
simplified-model benchmarks. An ablation study that removes the detailed
extra-jet kinematics reduces the expected significance by 53--71\%,
indicating that extra-jet topology carries substantial discriminating power
in the hadronic mono-$Z$ channel. These results are projected sensitivities
(no unblinding performed); the limitations and reproducibility of the study
are discussed in Sections~\ref{sec:limitations} and~\ref{sec:reproducibility}.
\end{abstract}
\section{Introduction}

The existence of dark matter (DM) is inferred from a wide range of
independent cosmological and astrophysical observations, yet its particle
nature remains one of the most important open questions in fundamental
physics. Collider experiments offer a complementary route to direct and
indirect detection: DM may be pair-produced in high-energy proton--proton
collisions, and the resulting momentum imbalance, quantified by the
missing transverse momentum (\MET), carries information about the DM
couplings to Standard Model (SM) particles \cite{abercrombie_dm_forum}.
Among the so-called mono-$X$ signatures, the mono-$Z$ topology---a $Z$
boson produced in association with a pair of stable DM particles---provides
a distinctive handle on simplified models with $s$-channel mediators that
radiate an on-shell $Z$ boson \cite{backovic_schannel,neubert_monoz}. The
$Z$ boson can be reconstructed either through its leptonic decay
$Z\to\ell^{+}\ell^{-}$ or its hadronic decay $Z\to q\bar{q}$. Although the
leptonic channel offers a clean, low-background signature, the hadronic
channel benefits from the substantially larger branching fraction of the
$Z$ to jets and is complementary to the mono-jet searches that dominate the
early Run~2 DM programme.

In a preceding study \cite{rasineni_nsf_monoz}, we demonstrated a
density-based search for DM in the leptonic mono-$Z$ final state using CMS
Run~2015D open data. That analysis trained five Neural Spline Flows
(NSFs) \cite{durkan_nsf,papamakarios_flows}---two channel-specific SM
background flows and three mediator-specific DM signal flows---and used the
per-event log-likelihood ratio between the DM and SM densities as the test
statistic, followed by a simultaneous signal-region/validation-region
profile-likelihood fit. The leptonic analysis was limited by a residual
background-modelling discrepancy in the high-\MET{} tail, which inflated the
observed-to-expected limit ratio by a factor of roughly 7--12 and prevented a
clean interpretation of the fitted signal strength.

The present work extends this programme to the hadronic mono-$Z$ channel and
introduces a fundamentally different density-modelling strategy. Instead of
closed-form normalizing flows with analytic densities, we model the
background density with a conditional flow-matching continuous normalizing
flow (CFM-CNF) \cite{lipman_flow_matching,chen_neural_ode}. Whereas an NSF
evaluates the density in closed form through a composition of invertible
spline coupling layers, a CNF defines the density implicitly through a
learned time-dependent vector field integrated along a probability-flow ODE,
with the logarithmic density obtained from the divergence trace via
Hutchinson's stochastic estimator \cite{grathwohl_ffjord}. This
continuous-time formulation is trained with the simulation-free conditional
flow-matching objective \cite{lipman_flow_matching}, which regresses the
model onto the velocity of a straight-line probability path and avoids the
architectural constraints of invertible coupling layers. The switch from a
two-hypothesis likelihood-ratio score to a single background-density
discriminant also changes the statistical treatment: rather than fitting
signal-plus-background templates in a profile likelihood, we evaluate
per-event background negative log-likelihoods on a held-out validation split
and convert the resulting yields into an Asimov expected significance
\cite{cowan_asymptotic}.

The hadronic channel introduces additional methodological challenges that
motivate several of the technical choices in this paper. First, the
reconstruction of a hadronic $Z$ candidate from a dijet pair requires a
dijet-mass window and dijet-angular selection, and the presence of additional
jets (beyond the two $Z$-candidate jets) creates features that are
physically undefined when no such jet exists. In the leptonic analysis these
missing-object features were handled with binary jet-presence indicators and
zero-imputation \cite{rasineni_nsf_monoz}; here we instead preserve the
``does-not-exist'' state as an explicit out-of-range sentinel value, avoiding
the fabrication of a spurious point mass in the physical continuum. Second,
because the background density is trained on real selected HTMHT events, we
score only a persisted held-out validation split and reweight the scored
events back to the full selected population, eliminating the in-sample
scoring bias that would otherwise inflate the background yield. Third, since
the \textsc{Delphes} fast simulation does not emulate the CMS high-level
trigger, we apply an offline trigger proxy to the simulated signal before
scoring. Finally, to avoid the look-elsewhere bias of an unconstrained
argmax over a noisy discriminant scan, we report the working point that
maximises the expected significance subject to a minimum background yield of
$B\geq20$ events.

The analysis uses the CMS Run~2015D HTMHT MINIAOD open dataset
\cite{cms_htmht_2015d}, corresponding to a validated recorded integrated
luminosity of $2.256382381~\invfb$ \cite{cms_data_policy,cms_lumi_guide},
and follows the open-data research practices advocated by the CMS open-data
community \cite{lassila_cms_opendata,bellis_cms_dp}. Dark-matter signal
samples are generated with the \texttt{DMsimp\_s\_spin1} simplified-model
implementation of an $s$-channel spin-1 mediator
\cite{abercrombie_dm_forum,backovic_schannel} using a reproducible
\textsc{MadGraph5\_aMC@NLO} + \textsc{Pythia8} + \textsc{Delphes} toolchain
\cite{alwall_mg5amc,sjostrand_pythia82,delphes3}. Under the post-fix
baseline procedure the analysis yields expected significances of
$2.89\sigma$, $7.62\sigma$, and $7.41\sigma$ for three simplified-model
benchmarks, and an ablation study that removes detailed extra-jet kinematics
reduces the expected significance by 53--71\%, indicating that extra-jet
topology carries substantial discriminating power in the hadronic mono-$Z$
channel.

The remainder of this paper is organised as follows. Section~2 describes the
data and simulation, including the hadronic mono-$Z$ selection and the
signal-generation toolchain. Section~3 presents the conditional flow-matching
background density model, the feature contract and preprocessing, and the
training procedure. Section~4 summarises the validation studies, including
closure checks and the before/after comparison of the sentinel-imputation
fix. Section~5 documents the sensitivity methodology, and Section~6 reports
the expected-sensitivity results. Section~7 presents the extra-jet feature
ablation study, Section~8 discusses limitations, and Section~9 concludes.
\section{Data and simulation}
\label{sec:data}

\subsection{CMS Run 2015D HTMHT open data}

The analysis uses the CMS Run~2015D HTMHT MINIAOD open dataset
\cite{cms_htmht_2015d}, corresponding to a validated recorded integrated
luminosity of $2.256382381~\invfb$ for runs 256630--260627
\cite{cms_data_policy,cms_lumi_guide}. Events are read in chunks via
parallelised XRootD access with a custom feature-extraction pipeline built on
\texttt{uproot} and \texttt{awkward}; per-file processing counts are recorded
to verify completeness of the dataset traversal
(Appendix~\ref{app:filecounts}). In total 444 MINIAOD files containing
20{,}679{,}437 raw events were processed, of which 1{,}439{,}523 events
satisfy the hadronic mono-$Z$ selection described below, corresponding to a
selection efficiency of 6.96\%.

Events are accepted if they pass any of the five unprescaled HLT trigger
paths listed in Appendix~\ref{app:hlt}, which comprise $H_T$-based,
\MET-based, and mono-jet trigger paths and accept 91.3\% of the raw events.
The offline hadronic mono-$Z$ selection requires at least two jets with
$p_T > 30~\GeV$ and $|\eta| < 2.4$. The two leading jets define the
hadronic-$Z$ candidate and must satisfy $70 < m_{jj} < 110~\GeV$ and
$\Delta R_{jj} < 2.0$. A b-tag veto is applied to additional jets beyond the
two $Z$-candidate jets. The selected events are stored in a single ROOT tree
with a fixed 41-branch feature schema (kinematic, angular, and recoil
observables; see Appendix~\ref{app:features}) and are used for the
background-density training and the expected-sensitivity calculation
described below.

The use of CMS open data in research brings specific challenges that have
been documented by the CMS open-data team \cite{lassila_cms_opendata}.
In particular, the authors note that the complexity of the CMS software
(CMSSW) and the intrinsic complexity of the data format pose a significant
hurdle for external users, and they emphasise the importance of providing
full analysis benchmarks and simplified data formats such as MiniAOD and
NanoAOD to improve usability \cite{lassila_cms_opendata}. The present
analysis addresses these challenges by reading the HTMHT MiniAOD files
directly with \texttt{uproot} and \texttt{awkward}, avoiding the need to
run the full CMSSW analysis chain, and by recording per-file processing
counts to verify the completeness of the dataset traversal. The trigger
decision is reconstructed from the \texttt{filterLabels\_} branch using a
custom decoder, following the guidance of the open-data documentation for
2015 MiniAOD files whose \texttt{pathNames\_} payload is empty
\cite{lassila_cms_opendata}. The luminosity information is taken from the
validated Run~2015D recorded value, following the open-data luminosity
guidance \cite{cms_lumi_guide}.

\subsection{Signal generation}

Dark-matter signal samples are generated with \textsc{MadGraph5\_aMC@NLO},
showered and hadronised with \textsc{Pythia8}, and passed through the CMS
card of a parameterised \textsc{Delphes} detector simulation
\cite{alwall_mg5amc,sjostrand_pythia82,delphes3}, with jets reconstructed
using the anti-$k_T$ algorithm \cite{cacciari_antikt}. The anti-$k_T$
algorithm was chosen for its soft-resilient jet boundaries: as shown in the
original paper \cite{cacciari_antikt}, its jets are conical with radius $R$
for isolated hard particles, and their boundaries are insensitive to soft
radiation while remaining flexible to hard substructure. This property
suppresses the back-reaction and area fluctuations that would otherwise
smear the jet momenta entering the hadronic-$Z$ reconstruction, and it is
the algorithm adopted by the CMS and ATLAS collaborations for Run~2
analyses. Dark-matter production is modelled with the
\texttt{DMsimp\_s\_spin1} simplified-model UFO implementation of an
$s$-channel spin-1 mediator coupling the dark and visible sectors
\cite{abercrombie_dm_forum,backovic_schannel}, for the process
$pp \to Z\bar{\chi}\chi$ with $Z \to jj$. Three benchmark points,
listed in Table~\ref{tab:signal-points}, are retained for the final
analysis; each is generated with 5{,}000 events and a fixed MadGraph5
run-card seed (110020, 150200, and 101500, respectively) to ensure
reproducibility, and the leading-order cross sections are taken from the
generated LHE event samples. A fourth, light-scalar-mediator benchmark was
generated but is excluded: the hadronic-$Z$ reconstruction does not recover
a usable dijet resonance for this point (the median closest-to-$m_Z$ dijet
mass remains at 145--178~\GeV, far from $m_Z \simeq 91.2~\GeV$).

The three retained benchmarks follow the benchmark framework recommended by
the ATLAS/CMS Dark Matter Forum \cite{abercrombie_dm_forum}, which
codified the $s$-channel spin-1 mediator models with vector and axial-vector
couplings together with a recommended grid in the
$(m_\chi, M_{\mathrm{med}})$ plane for early Run~2 searches. In
particular, the axial points $(m_\chi, M_{\mathrm{med}}) = (10, 20)~\GeV$
and $(50, 200)~\GeV$ lie in the threshold and on-shell regions of the
Forum's parameter grid, respectively, while the vector point
$(1, 500)~\GeV$ is an on-shell benchmark with a heavy mediator. The Forum's
parameter scans further showed that the shapes of the \MET{} and boson
$p_T$ distributions are insensitive to the coupling choices $g_q$ and
$g_\chi$ within the scanned range, with only the production cross section
changing; the kinematic emphasis of the present study on the hadronic-$Z$
reconstruction and extra-jet topology is therefore expected to hold across
the recommended coupling choices. Our use of the \texttt{DMsimp}
implementation follows the Forum's recommendation of a standardised
generator implementation for these models, at leading-order accuracy for
the $V + \MET$ final states \cite{backovic_schannel,abercrombie_dm_forum},
which is appropriate for a projected-sensitivity study of this type.

The leading-order cross sections used for signal normalisation should be
interpreted in light of the higher-order QCD predictions for these
s-channel mediator models \cite{backovic_schannel}. The DMsimp
implementation was validated and made available at next-to-leading order
(NLO) in QCD within the FeynRules/MadGraph5\_aMC@NLO framework
\cite{backovic_schannel}, and the NLO corrections to DM pair production in
association with jets were found to be significant for light mediators and
light DM, with K factors up to $\sim$1.8 for $(m_Y, m_X) = (10, 1)~\GeV$,
decreasing to $K \sim 1.1$--1.2 for heavy mediators and heavy DM
\cite{backovic_schannel}. For the benchmark points retained here, the
axial $(10, 20)~\GeV$ and $(50, 200)~\GeV$ points and the vector
$(1, 500)~\GeV$ point, the NLO corrections are expected to be moderate
($K \sim 1.1$--1.4), and the leading-order cross sections quoted in
Table~\ref{tab:signal-points} should be regarded as lower estimates of the
true production rates. The NLO study also showed that the shapes of the
\MET{} and jet $p_T$ distributions are largely preserved at NLO, with the
largest corrections in the low-\MET{} region \cite{backovic_schannel},
which is consistent with the kinematic emphasis of the present analysis on
the hadronic-$Z$ reconstruction and extra-jet topology rather than on the
absolute signal normalisation.

\begin{table*}[t]
  \centering
  \small
  \caption{Retained dark-matter signal benchmarks. \texttt{DMsimp\_s\_spin1}
  is an $s$-channel spin-1 mediator; $m_\chi$ and $m_{\mathrm{med}}$ are the
  dark-matter and mediator masses. Cross sections are leading-order
  MadGraph5 values.}
  \label{tab:signal-points}
  \setlength{\tabcolsep}{4pt}    
  \begin{tabular*}{0.9\textwidth}{@{\extracolsep{\fill}}L{4.8cm}llrr@{}}
    \toprule
    Signal point & Mediator & $(m_\chi,\,m_{\mathrm{med}})$ [\GeV] & $\sigma$ [pb] & $N_{\mathrm{sel}}$ \\
    \midrule
    \texttt{axial\_mx10\_mv20}  & axial  & $(10, 20)$  & 1.204 & 116 \\
    \texttt{axial\_mx50\_mv200} & axial  & $(50, 200)$ & 1.050 & 487 \\
    \texttt{vector\_mx1\_mv500} & vector & $(1, 500)$  & 0.544 & 848 \\
    \bottomrule
  \end{tabular*}
\end{table*}

After the offline hadronic mono-$Z$ selection, 116, 487, and 848 events
remain for the three benchmarks, respectively. Because \textsc{Delphes} does
not emulate the CMS high-level trigger, an offline trigger proxy
reconstructing the logic of the five HLT paths used on data
(Appendix~\ref{app:hlt}; \S\ref{sec:trigger-proxy}) is applied to the
simulated samples before scoring; all events surviving the offline selection
also satisfy the proxy, i.e.\ the trigger-proxy efficiency is 100\% for each
point.

\subsection{Z-candidate reconstruction and feature contract}

For the data sample the hadronic-$Z$ candidate is formed from the two leading
jets, while for simulated signal the candidate is chosen as the dijet pair
closest to $m_Z$ among the leading eight jets, consistent with the
signal-generation validation. Both choices use the same jet $p_T$ and
$\eta$ thresholds and produce a common 41-branch feature schema, locked via
the extractor schema contract so that data and signal share identical
feature definitions. Undefined angular observables that arise when no
additional jet exists (for example the
$\Delta\phi(\MET,\mathrm{jet}_{1,2})$ and extra-jet $\eta$ features) are
preserved as explicit out-of-range sentinel values rather than imputed into
the physical continuum; this handling and the final training feature set are
described in the methods section.

\section{Methods}

This section describes the background-density model, the preprocessing of
the extracted features, the training procedure and hyperparameters, and the
evaluation of per-event densities used in the sensitivity calculation.

\subsection{Background density model}

The background density is modelled with a conditional flow-matching
continuous normalizing flow (CFM-CNF) \cite{lipman_flow_matching}. The
``conditional'' refers to the standard conditional flow-matching
construction, in which the flow is conditioned on its target data point in
the training objective; the model is not conditioned on auxiliary
covariates. A continuous normalizing flow \cite{chen_neural_ode,
papamakarios_flows} maps a base Gaussian $\mathbf{z}_0 \sim
\mathcal{N}(0, I_D)$ onto standardised feature vectors $\mathbf{x}\in
\mathbb{R}^D$ by integrating a learned time-conditioned vector field
$v_\theta: \mathbb{R}^D\times[0,1]\to\mathbb{R}^D$,
\begin{equation}
  \frac{\mathrm{d}\mathbf{z}}{\mathrm{d}t}
  = v_\theta\!\left(\mathbf{z}(t), t\right), \qquad
  \mathbf{z}(1)=\mathbf{x}, \quad \mathbf{z}(0)=\mathbf{z}_0,
  \label{eq:pflow}
\end{equation}
and the per-event density is obtained from the change of variables along
this ODE. In contrast to closed-form normalizing-flow architectures such as
the neural spline flows used in previous work on this channel
\cite{durkan_nsf,rasineni_nsf_monoz}, the CNF density is evaluated by
numerically integrating Eq.~\eqref{eq:pflow}, which is computationally more
expensive but retains a fully continuous-time density model.

The choice of a CNF over the NSF architecture used in the leptonic
predecessor \cite{rasineni_nsf_monoz} reflects several methodological
differences. An NSF evaluates the density in closed form through a
composition of invertible rational-quadratic spline coupling layers, which
is fast and exact but constrains the model to a fixed, layer-wise
transformation. A CNF instead defines the density implicitly through a
learned time-dependent vector field, which is more flexible but requires
numerical ODE integration and a stochastic divergence-trace estimate for
each density evaluation. In the leptonic analysis the NSF score was
constructed as a two-hypothesis log-likelihood ratio between a DM flow and
a channel-specific SM flow, and the final interpretation used a
simultaneous signal-region/validation-region profile-likelihood fit with
per-channel normalisation nuisances. Here we instead train a single
background-density model and use the per-event negative log-likelihood as a
one-sided discriminant, converting the yields above a threshold into an
Asimov expected significance \cite{cowan_asymptotic}. This simplification
is appropriate for a projected-sensitivity study in which no unblinding is
performed, and it avoids the background-modelling residual that dominated
the leptonic analysis's high-\MET{} tail \cite{rasineni_nsf_monoz}.

The continuous-time formulation adopted here follows the continuous
normalizing flow construction introduced in the neural-ODE framework
\cite{chen_neural_ode}, which showed that the change in log density under a
continuous transformation is governed by the trace of the vector-field
Jacobian rather than the log determinant of a finite transformation. This
replaces the cubic-cost determinant computation of discrete flows with a
trace that can be estimated in linear time, at the price of numerical ODE
integration for both the state and the accumulated log density
\cite{chen_neural_ode}. The normalizing-flow review of Papamakarios et al.
\cite{papamakarios_flows} classifies flow architectures into finite
compositions of invertible transformations and continuous-time
(infinitesimal) flows, and notes that the two families differ
fundamentally in how they trade off expressivity, invertibility, and
computational cost. The present work sits in the continuous-time family:
unlike the finite-composition NSF used in the leptonic predecessor
\cite{durkan_nsf,rasineni_nsf_monoz}, the CNF does not require a
partitioning or ordering of the feature dimensions and can be trained by
maximum likelihood without the architectural constraints of coupling or
autoregressive layers \cite{chen_neural_ode,papamakarios_flows}. The
review also emphasises that the choice between evaluating the forward or
inverse transformation is not a crucial implementation decision for
continuous-time flows, since both directions have the same computational
complexity \cite{papamakarios_flows}; this symmetry is exploited here by
evaluating the background density directly through the probability-flow
ODE.

The two building blocks of the leptonic predecessor are also worth
contrasting explicitly. The NSF architecture of Durkan et al.
\cite{durkan_nsf} parameterises the elementwise transformations in coupling
or autoregressive layers as monotonic rational-quadratic splines, which are
analytically invertible and have a closed-form Jacobian determinant. This
provides a powerful but finite-composition flow whose flexibility is
bounded by the number and width of the coupling layers. The present work
instead follows the FFJORD construction of Grathwohl et al.
\cite{grathwohl_ffjord}, which showed that the continuous-time
instantaneous change of variables of Chen et al. \cite{chen_neural_ode}
can be scaled to high-dimensional problems by estimating the Jacobian trace
with Hutchinson's stochastic estimator \cite{grathwohl_ffjord}. Because the
trace estimator has a cost of $\mathcal{O}(D)$ per evaluation rather than
the $\mathcal{O}(D^2)$ cost of an exact trace or the $\mathcal{O}(D^3)$
cost of a Jacobian determinant, the CFM-CNF is free to use an unrestricted
network architecture for the vector field. This is the same design choice
adopted here: the residual-MLP vector field described below is not
constrained by coupling-layer or autoregressive structure, and the density
is evaluated with a single Rademacher projection per ODE evaluation, as in
FFJORD \cite{grathwohl_ffjord}. The learned residual-MLP vector field can
therefore capture correlations between all feature dimensions without the
partitioning or ordering requirements of the NSF coupling layers used in
the leptonic analysis \cite{durkan_nsf,rasineni_nsf_monoz}.

The vector field is trained with the simulation-free conditional
flow-matching objective \cite{lipman_flow_matching}, which regresses the
model onto the velocity of the straight-line probability path between a
Gaussian sample $\mathbf{x}_0$ and a data point $\mathbf{x}_1$,
\begin{equation}
  \begin{split}
  \mathcal{L}_{\mathrm{FM}}(\theta)
  &= \mathbb{E}\!\left[
  \left\lVert
  v_\theta(\mathbf{x}_t,t)-(\mathbf{x}_1-\mathbf{x}_0)
  \right\rVert^2
  \right],\\
  &\quad
  t\sim U[0,1],\;
  \mathbf{x}_0\sim\mathcal{N}(0,I),\;
  \mathbf{x}_1\sim p,\\
  &\quad
  \mathbf{x}_t=(1-t)\mathbf{x}_0+t\mathbf{x}_1.
  \end{split}
  \label{eq:fmloss}
\end{equation}
so the training loss is a mean-squared-error regression on target
velocities, evaluated per mini-batch. At inference, the logarithmic density
is
\begin{equation}
  \log p(\mathbf{x}) = \log p_0(\mathbf{z}_0) - \int_0^1 \operatorname{Tr}\!\left( \frac{\partial v_\theta}{\partial \mathbf{z}} \right) \mathrm{d}t,
  \label{eq:logprob}
\end{equation}
where $p_0$ is the unit Gaussian density and the divergence trace is
estimated with Hutchinson's stochastic estimator using a single Rademacher
projection per ODE evaluation \cite{grathwohl_ffjord}. The probability-flow
ODE is solved with a fixed-step RK4 solver (step size 0.05, absolute and
relative tolerances $10^{-3}$); density values are accumulated in batches of
16 events.

\subsection{Feature contract and preprocessing}

From the shared 41-branch data/signal schema, 32 features are used for
training. The remaining nine columns are excluded because they are either
audit bookkeeping columns (\texttt{source\_file\_idx}, \texttt{source\_entry}),
trigger or veto flags (\texttt{trigger\_pass}, \texttt{btag\_pass},
\texttt{n\_bjets}), or columns whose data/MC semantics are known to differ
(\texttt{max\_btag\_csvv2} and \texttt{btag\_wp\_medium}, because Delphes
b-tagging is a binary flag rather than a continuous CSVv2 discriminator; and
\texttt{n\_vertices} and \texttt{rho}, because the signal simulation does not
include a pileup overlay while the data carries real pileup conditions).

Undefined extra-jet angular observables---\texttt{jet1\_eta},
\texttt{jet2\_eta}, \texttt{dphi\_met\_jet1}, and \texttt{dphi\_met\_jet2},
which are NaN when the corresponding extra jet does not exist---are mapped
to an explicit out-of-range sentinel value of $-999$ before
standardisation, rather than median-imputed into the physical continuum,
which would fabricate a spurious point mass. This sentinel handling is
applied identically to background, validation, and signal samples. Any
remaining non-finite values are filled with training-set medians, and all
features are standardised using the training-set mean and standard
deviation; the saved preprocessor is applied unchanged to all downstream
samples. Events are split 80/20 into training and validation subsets using
a fixed random seed (20260730), and the split indices are persisted so that
all density and sensitivity evaluations use the exact same held-out
validation split.

\subsection{Architecture and training}

The vector field is a residual MLP: a sinusoidal time embedding of
dimension 32 is concatenated to the input and mapped by an input layer to a
hidden width of 192; the result passes through four residual blocks, each
consisting of \texttt{LayerNorm}, \texttt{SiLU}, a 192-wide linear layer,
\texttt{SiLU}, \texttt{Dropout(0.05)}, and a second 192-wide linear layer,
followed by a residual connection; the output head is
\texttt{LayerNorm}, \texttt{SiLU}, and a final linear layer back to the
feature dimension $D$.

Training minimises Eq.~\eqref{eq:fmloss} with \texttt{AdamW}
($\texttt{lr}=2\times10^{-4}$, weight decay $10^{-5}$), a batch size of
2048, and gradient clipping at norm 5.0, for at most 500 epochs. Because the
flow-matching loss is not a direct density objective, convergence is
monitored every five epochs on the held-out validation negative
log-likelihood from Eq.~\eqref{eq:logprob}, with early stopping after six
consecutive checks without improvement. The baseline fit used
$D=32$ features on 1{,}439{,}523 selected background events (1{,}151{,}618
training, 287{,}905 validation) and its best checkpoint is at epoch 245
(validation NLL $\approx -64.97$). The ablation fit is identical except that
the six extra-jet features ($\texttt{jet1\_pt}$,
$\texttt{jet1\_eta}$, $\texttt{jet2\_pt}$, $\texttt{jet2\_eta}$,
$\texttt{dphi\_met\_jet1}$, $\texttt{dphi\_met\_jet2}$) are removed,
leaving $D=26$ features; its best checkpoint is at epoch 165 (validation NLL
$\approx -43.63$). The smaller validation-NLL magnitude of the ablation fit
is expected from the lower-dimensional density and does not indicate worse
training.

\subsection{Inference and sensitivity evaluation}

Per-event negative log-likelihoods (NLL) from Eq.~\eqref{eq:logprob} serve
as the discriminant between background and signal. Sensible comparison
requires the background NLL distribution to be evaluated on events not used
for training; we therefore score only the persisted held-out validation
split and reweight the scored events back to the full selected background
population, as detailed in Section~\ref{sec:sens-methodology}. Signal
samples are scored only after the offline trigger proxy described in
Section~\ref{sec:trigger-proxy}. The expected-significance calculation,
the minimum-background-yield working point, and the diagnostic closure
studies are described in Sections~\ref{sec:sens-methodology} and
\ref{sec:validation}.

\section{Validation}
\label{sec:validation}

This section summarises the validation studies used to verify the background
density model and to demonstrate that the sentinel-imputation fix removed
the dominant imputation artifact without introducing regressions in other
features.

\subsection{Closure plots (representative set)}

Closure assesses how faithfully the trained flow reproduces the held-out
validation data distribution, which is never used for training. We compare
histograms of the real validation events with samples drawn from the fitted
flow after inverse standardisation, using the same binning in both cases.
Representative panels are shown in Figure~\ref{fig:closure-representative},
and the full per-feature closure set is provided in
Appendix~\ref{app:closures}.

The hadronic-$Z$ mass panel (top left of
Figure~\ref{fig:closure-representative}) shows that the flow reproduces the
$Z$-peak structure inside the $70$--$110~\GeV$ window and tracks the
validation data across the full displayed range. A localised excess is
visible in the flow sample near $m_{jj}\approx 91.6~\GeV$ (density
$\sim\!0.038$ versus $\sim\!0.024$ in the held-out data, a $\sim$50\%
discrepancy in that bin), consistent with the continuous-density model
slightly over-concentrating probability at the sharp $Z$ pole; this is a
sub-bin-shape effect rather than an order-of-magnitude mismatch. The \MET{}
panel (top right) and the $\Delta R_{Zjj}$ panel (bottom right) confirm good
agreement for the missing-transverse-momentum and dijet-angular observables
that drive the mono-$Z$ discrimination, while the $H_T$ panel (bottom left)
verifies closure for the overall event scale. In all panels the model
histogram follows the data within statistical fluctuations of the validation
sample, with no order-of-magnitude mismatches.

\begin{figure*}[t]
  \centering
  \includegraphics[width=0.48\textwidth]{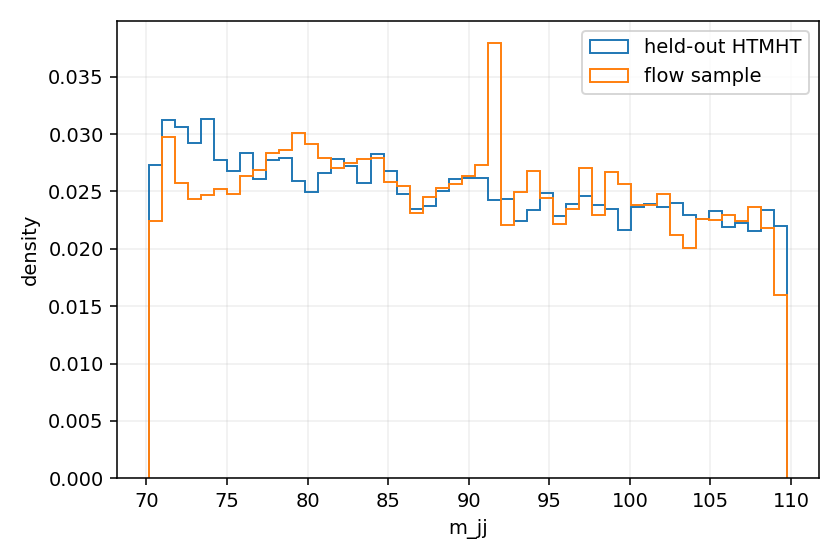}\hfill
  \includegraphics[width=0.48\textwidth]{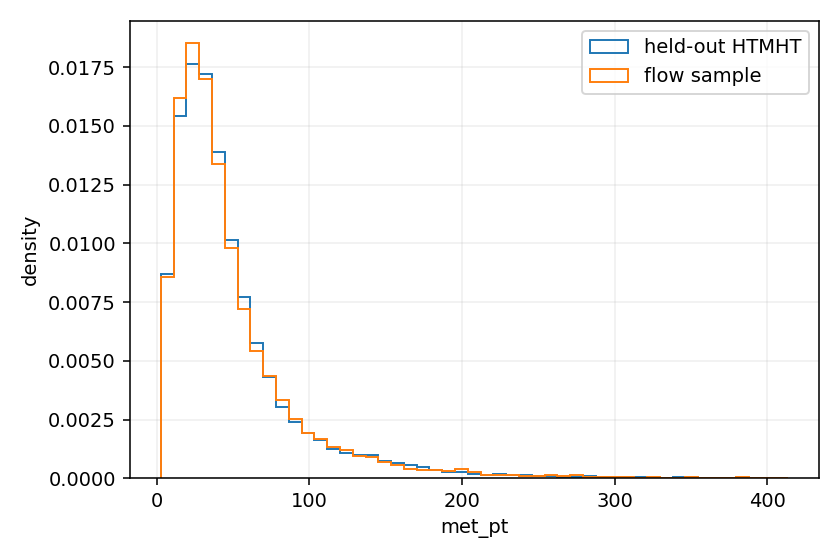}\\[6pt]
  \includegraphics[width=0.48\textwidth]{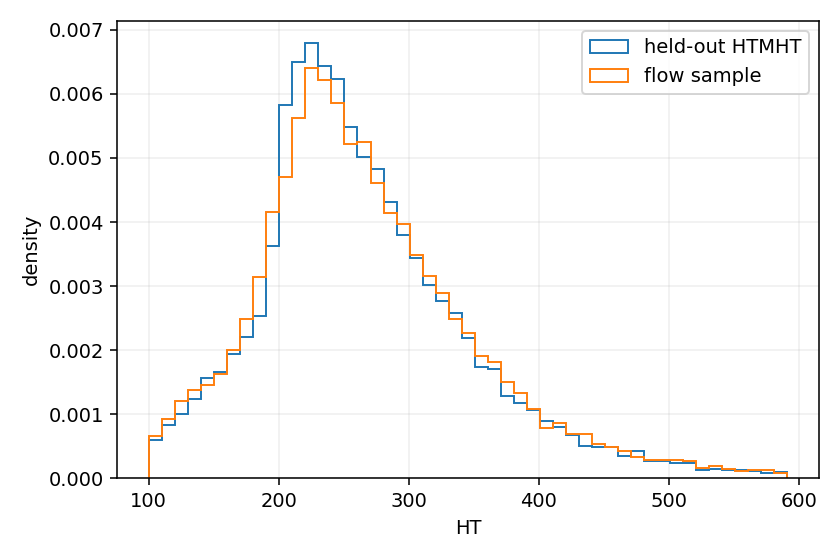}\hfill
  \includegraphics[width=0.48\textwidth]{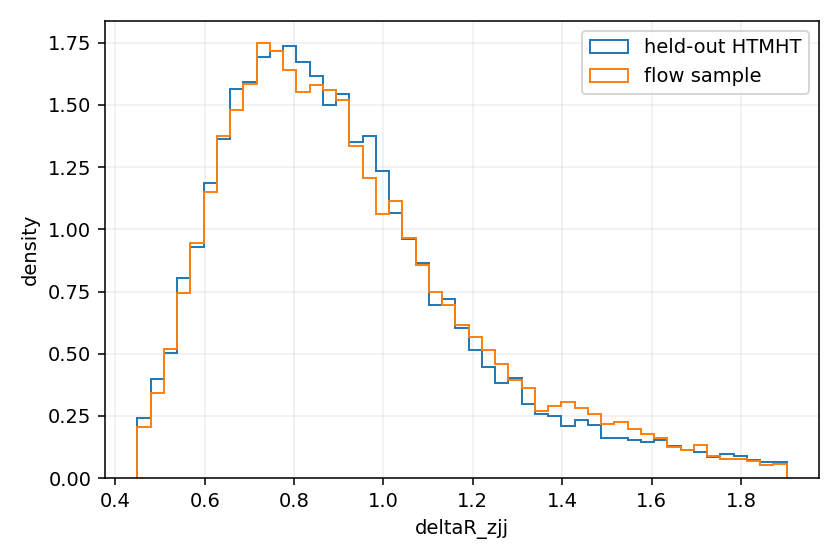}
  \caption{Representative closure plots comparing held-out validation data
  (filled histogram) to samples drawn from the fitted flow (step histogram).
  Top left: hadronic-$Z$ mass $m_{jj}$. Top right: $\MET$ distribution.
  Bottom left: event scalar $H_T$. Bottom right: $\Delta R_{Zjj}$. The full
  per-feature closure set is provided in Appendix~\ref{app:closures}.}
  \label{fig:closure-representative}
\end{figure*}

\begin{figure*}[t]
  \centering
  \includegraphics[width=0.48\textwidth]{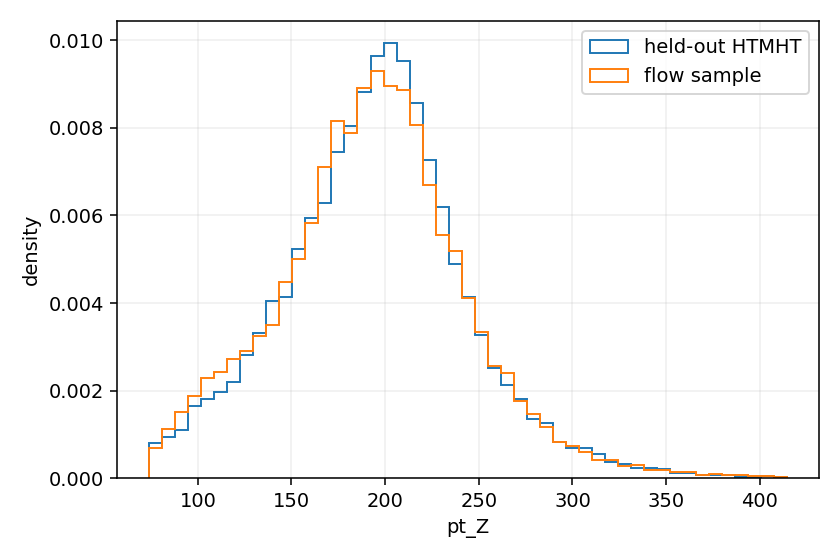}\hfill
  \includegraphics[width=0.48\textwidth]{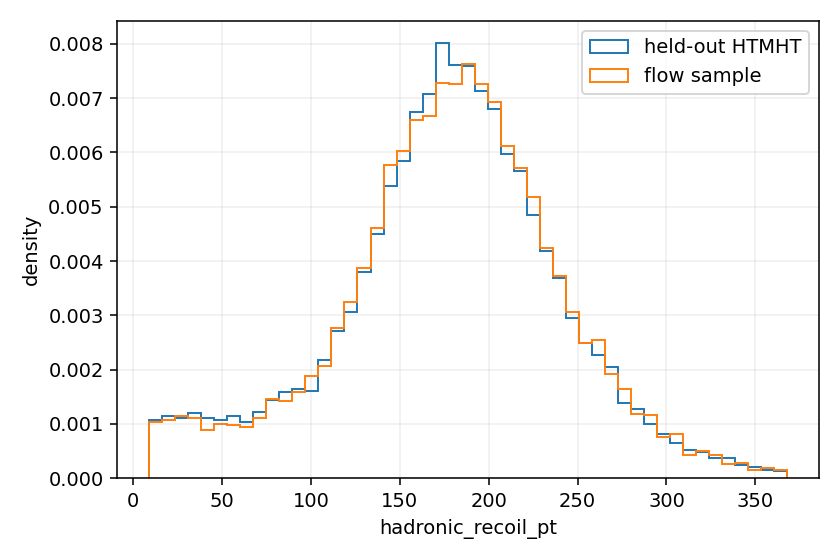}\\[6pt]
  \includegraphics[width=0.48\textwidth]{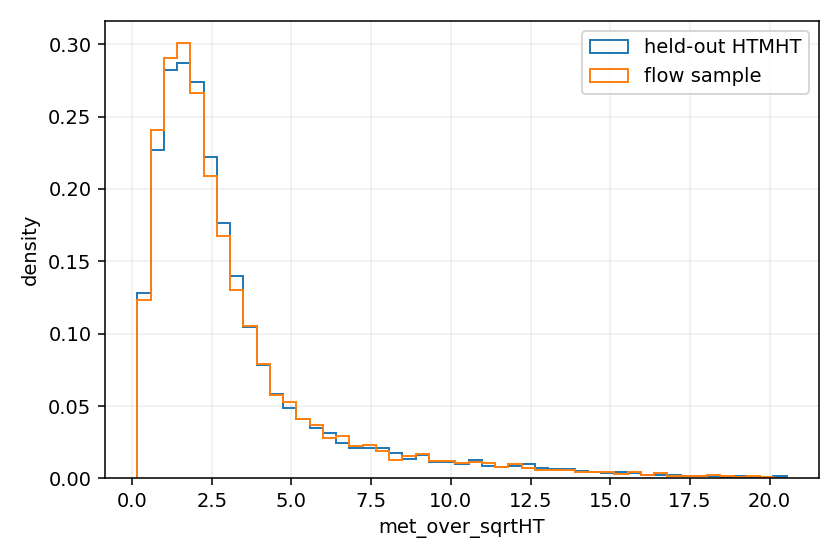}\hfill
  \includegraphics[width=0.48\textwidth]{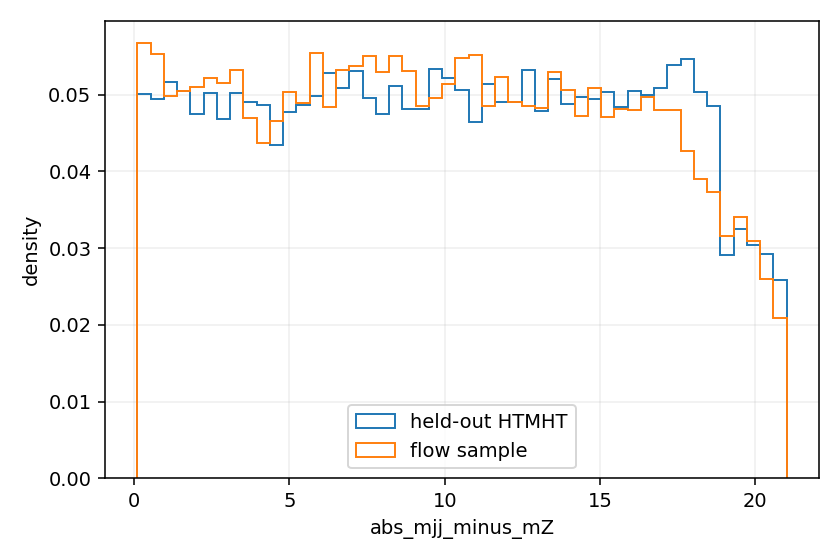}
  \caption{Additional closure panels for kinematic observables entering the
  density score. Top left: reconstructed $Z$ transverse momentum $p_{T,Z}$.
  Top right: hadronic recoil $p_T$. Bottom left: $\MET/\sqrt{H_T}$. Bottom
  right: $|m_{jj}-m_Z|$. The full per-feature closure set is provided in
  Appendix~\ref{app:closures}.}
  \label{fig:closure-kinematics}
\end{figure*}

Figure~\ref{fig:closure-kinematics} shows additional closure panels for the
reconstructed $Z$ transverse momentum, the hadronic recoil $p_T$, the
$\MET/\sqrt{H_T}$ ratio, and $|m_{jj}-m_Z|$. Good closure for these derived
observables indicates that the flow captures not only the individual jet
distributions but also the correlations encoded in the $Z$ kinematics and
the recoil system, which are the primary inputs to the per-event density
score used in Section~\ref{sec:sens-methodology}.

\subsection{Before/after imputation-fix comparison}

Auditing the original preprocessing revealed that four undefined extra-jet
angular features---\texttt{jet1\_eta}, \texttt{jet2\_eta},
\texttt{dphi\_met\_jet1}, \texttt{dphi\_met\_jet2}---were median-imputed when
no extra jet exists. Because these observables are physically undefined for
events without additional jets, the median fill collapsed the
``no-extra-jet'' state into a spurious point mass inside the physical
continuum, producing artificial spike-blending artifacts in the closure
plots. The post-fix pipeline maps these undefined values to an explicit
out-of-range sentinel value of $-999$
($\texttt{nan\_sentinel\_value} = -999.0$), applied identically to
background, validation, and signal samples before standardisation.

\begin{figure*}[t]
  \centering
  \includegraphics[width=0.48\textwidth]{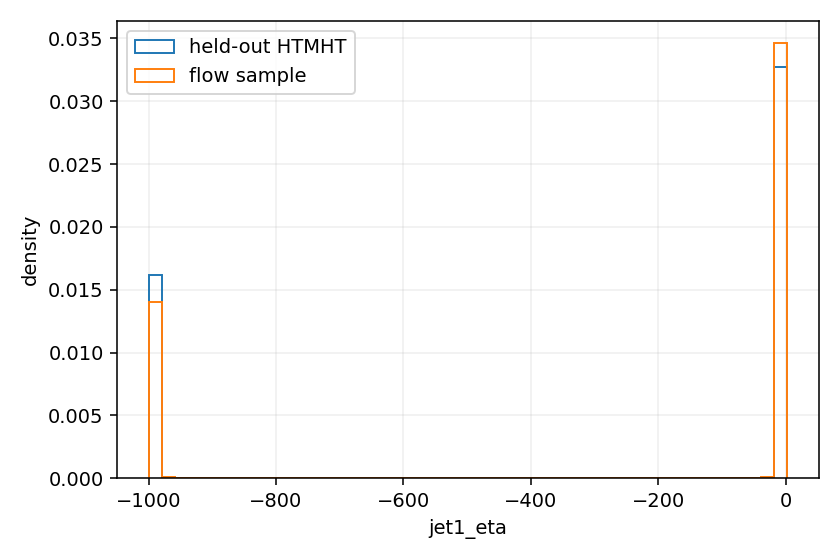}\hfill
  \includegraphics[width=0.48\textwidth]{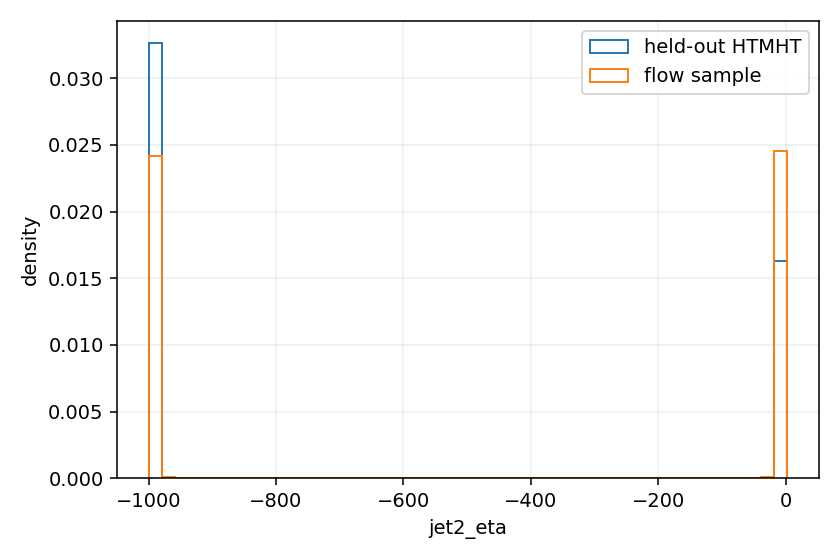}\\[6pt]
  \includegraphics[width=0.48\textwidth]{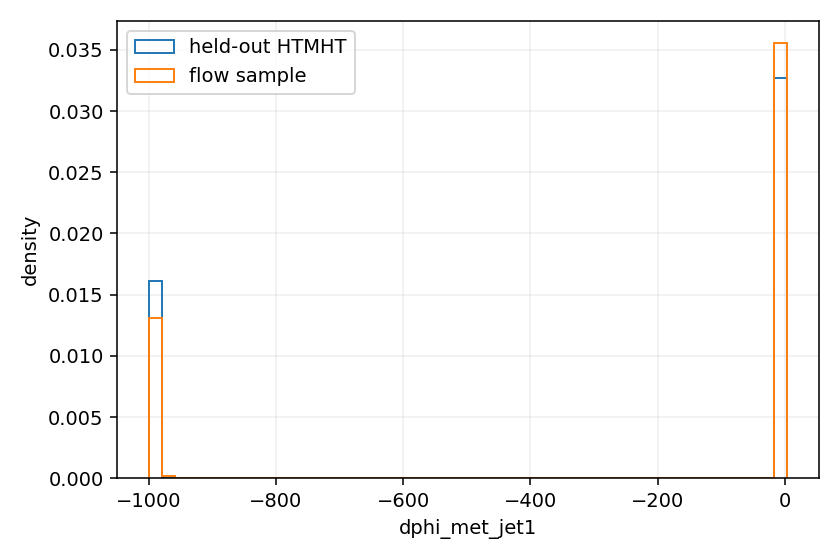}\hfill
  \includegraphics[width=0.48\textwidth]{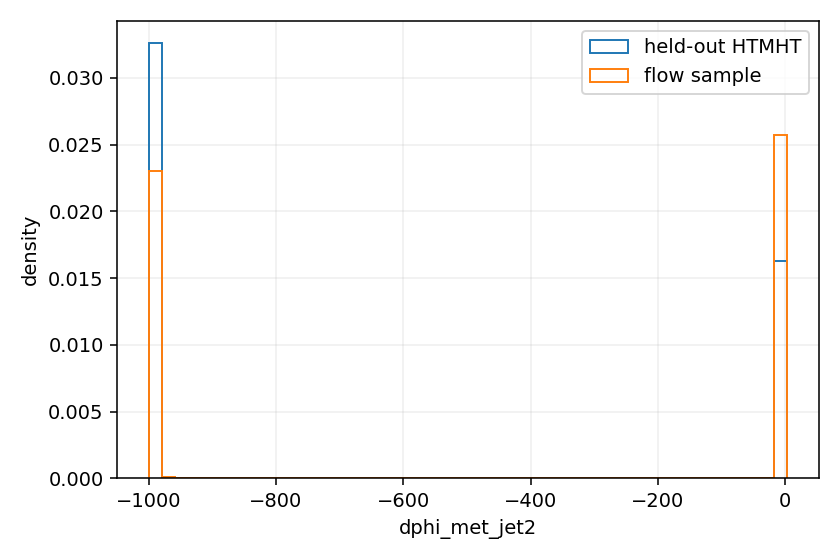}
  \caption{Post-fix closure panels for the four undefined extra-jet angular
  features. Top left: \texttt{jet1\_eta}. Top right: \texttt{jet2\_eta}.
  Bottom left: \texttt{dphi\_met\_jet1}. Bottom right: \texttt{dphi\_met\_jet2}.
  Each panel shows a distinct sentinel bar at $-999$ (the ``no extra jet''
  state) plus the physical continuum, both reproduced by the flow.}
  \label{fig:imputation-comparison}
\end{figure*}

Figure~\ref{fig:imputation-comparison} shows the post-fix closure panels for
all four fixed features. Each panel now exhibits two cleanly separated
populations: a sharp isolated bar at the sentinel value (the
``does-not-exist'' state) and a smooth histogram over physical values, both
reproduced by the fitted flow with reasonable closure. In particular, the
\texttt{dphi\_met\_jet2} panel previously showed a flow spike at density
$\sim\!10.9$ against a data value of $\sim\!5.45$---an order-of-magnitude
mismatch from the median fill---and now shows a substantially reduced but
still visible sentinel-bar mismatch (flow density $\sim\!0.023$ versus data
$\sim\!0.033$, a $\sim$30\% residual discrepancy) together with reasonable
agreement over the physical $\Delta\phi$ range. A comparable residual
sentinel-bar mismatch is visible in the \texttt{jet2\_eta} panel, while
\texttt{jet1\_eta} and \texttt{dphi\_met\_jet1} show closer agreement.

\subsection{Training convergence and validation NLL}

\begin{figure*}[tbp]
  \centering
  \includegraphics[width=0.92\textwidth]{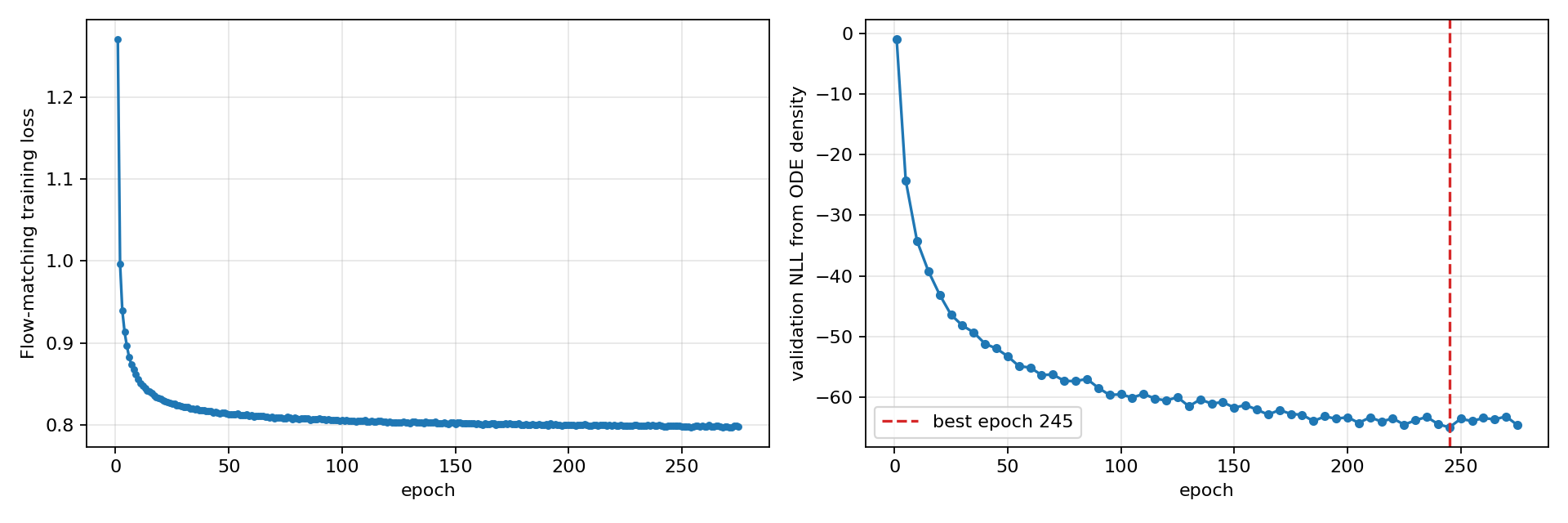}

  \vspace{0.8em}

  \includegraphics[width=0.92\textwidth]{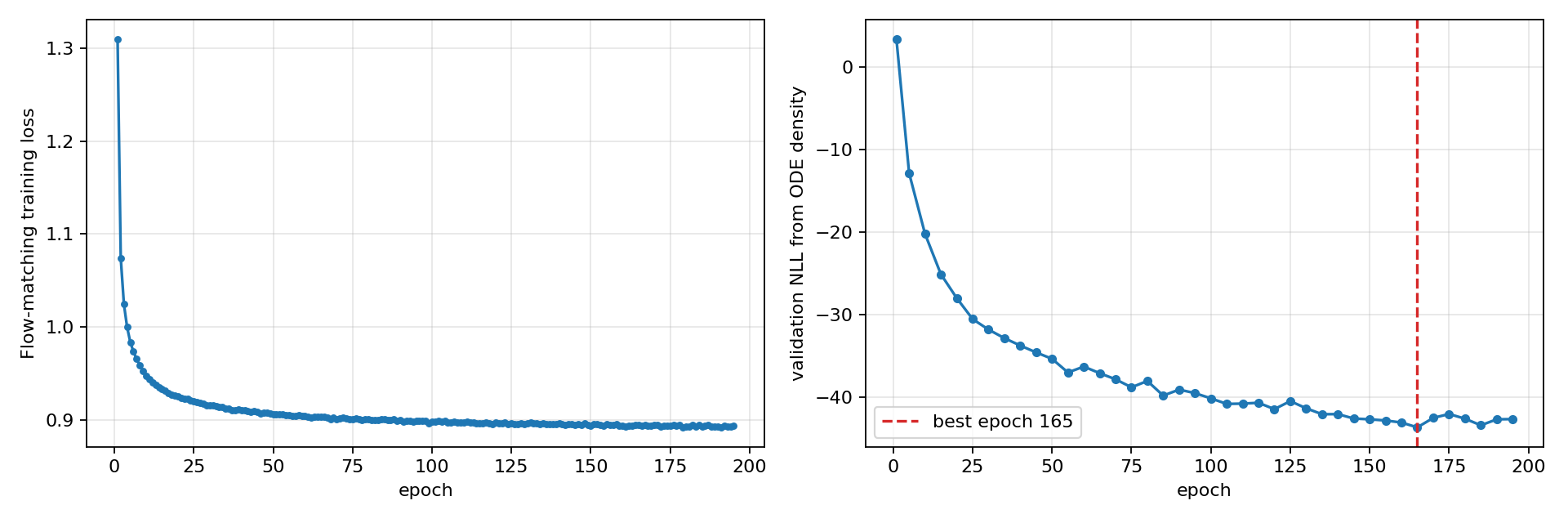}
  \caption{Validation negative-log-likelihood curves for the baseline (top)
  and extra-jet-drop ablation (bottom) fits. The red dashed line marks the
  best epoch selected by early stopping.}
  \label{fig:validation-nll}
\end{figure*}

Because the flow-matching loss is a regression objective rather than a
direct density loss, convergence is monitored on the held-out validation
negative log-likelihood obtained from the ODE density evaluation
(Eq.~\eqref{eq:logprob}). Figure~\ref{fig:validation-nll} shows the
validation-NLL curves for the baseline ($D=32$) and ablation ($D=26$) fits.
Both descend smoothly and plateau cleanly over the final $\sim$30--50
epochs, with no evidence that training stopped while the model was still
improving. The baseline selects its best checkpoint at epoch 245
(validation NLL $\approx -64.97$) and the ablation at epoch 165
(validation NLL $\approx -43.63$). The smaller magnitude of the ablation
value is expected for a lower-dimensional density and does not indicate
worse training; the closure checks above confirm that the ablation model
remains of comparable quality on its retained features.

\subsection{Known residual limitation: jet \texorpdfstring{$p_T$}{pT} delta-plus-continuum}

\begin{figure*}[t]
  \centering
  \includegraphics[width=0.48\textwidth]{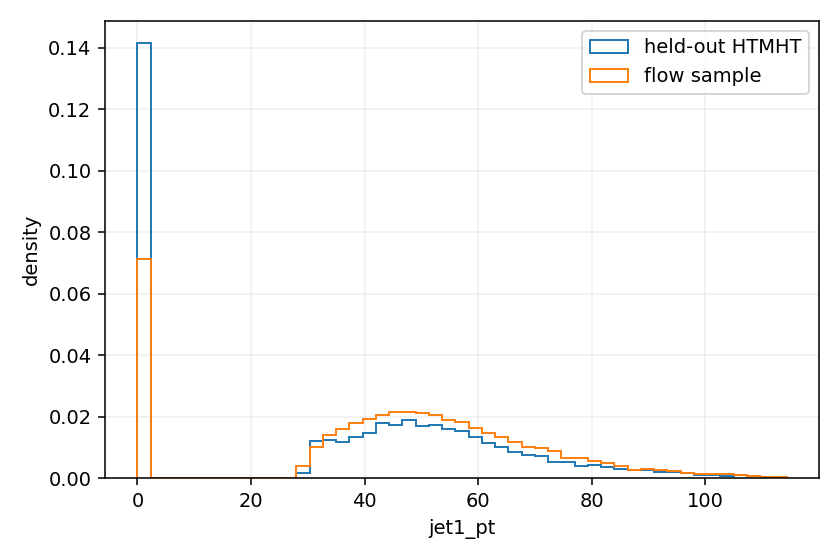}\hfill
  \includegraphics[width=0.48\textwidth]{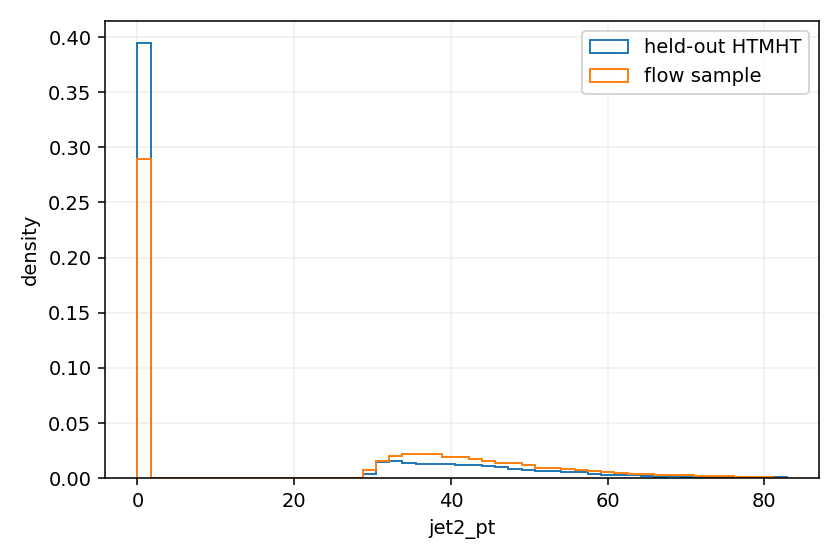}
  \caption{Closure panels for the extra-jet $p_T$ features \texttt{jet1\_pt}
  (left) and \texttt{jet2\_pt} (right), illustrating the residual
  delta-plus-continuum limitation: the flow smooths the physical $p_T=0$
  sentinel spike, undershooting its height while preserving the continuum.}
  \label{fig:pt-limitation}
\end{figure*}

A remaining methodological caveat is the inability of a continuous
normalizing flow to exactly represent a discrete point mass (the $p_T=0$
spike for absent extra jets) combined with a continuous distribution for
present jets within the same feature. The sentinel mapping resolves the
undefined-angular features, but the extra-jet $p_T$ features carry a
legitimate physical $p_T=0$ sentinel (zero-fill) alongside a continuum, and
the CNF necessarily smooths the point mass. As shown in
Figure~\ref{fig:pt-limitation}, the closure for \texttt{jet1\_pt} and
\texttt{jet2\_pt} retains a mild undershoot of the zero-spike height (for
example, flow density $\sim 0.071$ versus data $\sim 0.142$ for
\texttt{jet1\_pt}). This is a structural limitation of continuous-density
models for this feature type, not a training defect; it is stated explicitly
as a caveat in Section~\ref{sec:limitations}.

\section{Sensitivity methodology}
\label{sec:sens-methodology}

This section documents the statistical and procedural choices used to
convert the per-event background-density scores into the expected
significances reported in Section~\ref{sec:results}. The per-event negative
log-likelihood (NLL) from Eq.~\eqref{eq:logprob} serves as the discriminant:
signal-like events are expected to populate the high-NLL tail of the
background distribution, and a threshold on this score defines a counting
region in which the expected signal and background yields $S$ and $B$ are
evaluated.

\subsection{Asimov significance}

Expected significances are computed with the Asimov counting formula
\cite{cowan_asymptotic},
\begin{equation}
  Z_A = \sqrt{2\left[(S+B)\ln\left(1+\frac{S}{B}\right)-S\right]},
  \label{eq:asimov}
\end{equation}
where $S$ and $B$ are the weighted signal and background yields above the
chosen NLL threshold. This is a projected, expected significance: it is
evaluated from the simulated signal and the background-density model alone,
without comparing real data to a background prediction, and therefore
corresponds to the median significance that would be obtained in an
ensemble of background-only pseudo-experiments.

The use of the Asimov counting formula follows the asymptotic framework of
Cowan et al. \cite{cowan_asymptotic}, which derived closed-form sampling
distributions for profile-likelihood-ratio test statistics and introduced
the Asimov data set as a tool for estimating the median expected
significance without Monte Carlo simulation. In that framework, the Asimov
data set is constructed with all parameters equal to their expectation
values, and the median discovery significance for a signal hypothesis is
approximated by $\mathrm{med}[Z_0|\mu'] = \sqrt{q_{0,A}}$
\cite{cowan_asymptotic}. The counting formula of Eq.~\eqref{eq:asimov} is
the binomial-expansion limit of this result for a single-bin counting
experiment with negligible background uncertainty; it corresponds to the
median significance one would obtain from the profile-likelihood-ratio
statistic $q_0$ in the large-sample limit \cite{cowan_asymptotic}. The
full formal treatment with nuisance parameters would generalise
Eq.~\eqref{eq:asimov} to a profile likelihood over the background-density
model and the signal-yield normalisation, at the cost of substantially
greater computational complexity. For a projected-sensitivity study the
counting approximation is a well-motivated simplification, and the
$B\geq20$ working-point floor ensures that the Gaussian approximations
underlying the asymptotic formula remain reasonable, as recommended in
\cite{cowan_asymptotic}.

\subsection{Signal-yield normalization}

The expected signal yield above a threshold is obtained by normalising the
number of simulated signal events that pass the threshold to the integrated
luminosity. For each benchmark, the per-event weight is
\begin{equation}
  w_{\rm sig} = \frac{\sigma\, \mathcal{L}\, \epsilon_{\rm trig}}{N_{\rm gen}},
  \label{eq:signal-weight}
\end{equation}
where $\sigma$ is the leading-order cross section, $\mathcal{L}$ is the
integrated luminosity, $\epsilon_{\rm trig}$ is the signal-side trigger-proxy
efficiency, and $N_{\rm gen}$ is the number of generated signal events used
for normalisation. The signal yield above a threshold is then
$S = w_{\rm sig}\, N_{\rm sig}^{>\mathrm{cut}}$, where
$N_{\rm sig}^{>\mathrm{cut}}$ is the number of simulated signal events whose
NLL exceeds the threshold. The cross sections and generated-event counts are
taken from the signal-generation manifest (Table~\ref{tab:signal-points}),
and the luminosity is the validated Run~2015D recorded value of
$2.256382381~\invfb$ \cite{cms_data_policy,cms_lumi_guide}.

\subsection{Discriminant scan and working-point selection}

The NLL threshold is scanned over the combined set of background and signal
scores, and at each threshold the yields $S$ and $B$ are evaluated and
$Z_A$ is computed from Eq.~\eqref{eq:asimov}. During early audits we found
that reporting the unconstrained argmax of this scan can select thresholds
with vanishing background yields (raw counts as small as 3--12 events in the
full selected population), producing unstable and biased headline numbers
because the optimisation effectively performs a look-elsewhere search on
stochastic background fluctuations in the tail. To avoid this pitfall, the
primary reported working point is the best threshold satisfying a minimum
weighted background yield of $B\geq20$; the unconstrained optimum is still
recorded for diagnostic comparison but is not used as the headline result.
The $B\geq20$ floor ensures the reported $Z_A$ is evaluated at a working
point with sufficient background statistics that the Gaussian
approximations underlying the asymptotic formula are reasonable and that
the optimisation is not dominated by rare-tail sampling noise. As a
diagnostic, the relative Poisson uncertainty on the background yield,
$1/\sqrt{B}$, is also recorded at the selected working point.

\subsection{Held-out-only background scoring with reweighting}

To avoid in-sample scoring bias, the background NLL distribution is
evaluated only on the held-out validation split, which is never used for
training. The scored events are then reweighted back to the full selected
background population via
\begin{equation}
  w_{\rm pop} = \frac{N_{\rm selected}}{N_{\rm scored}},
  \label{eq:pop-weight}
\end{equation}
where $N_{\rm selected}=1{,}439{,}523$ is the post-selection population and
$N_{\rm scored}=287{,}905$ is the held-out validation count in the baseline
run. The background yield above a threshold is
$B = w_{\rm pop}\, N_{\rm bg}^{>\mathrm{cut}}$, where
$N_{\rm bg}^{>\mathrm{cut}}$ is the number of held-out background events
whose NLL exceeds the threshold. This provides a held-out estimate of the
full-population background yield while avoiding the positive bias introduced
by scoring training-set rows.

\subsection{Signal-side trigger proxy}
\label{sec:trigger-proxy}

Because Delphes does not emulate the CMS HLT turn-on, the signal samples
are processed with an offline trigger proxy approximating the five HLT
paths used on the recorded background. The adopted proxy applied to signal
is $H_T>350~\GeV$ OR $\MET>120~\GeV$ OR ($\MET>90~\GeV$ and leading jet
$p_T>80~\GeV$). After the hadronic mono-$Z$ offline selection, 116, 487, and
848 events remain for the axial 10/20, axial 50/200, and vector 1/500
benchmarks, respectively; all satisfy the proxy, so the trigger-proxy
efficiency is 100\% for each point and $\epsilon_{\rm trig}=1$ in
Eq.~\eqref{eq:signal-weight}. This proxy is conservative but does not
replace a full HLT efficiency measurement, which would require a data-driven
turn-on parameterisation.

\section{Results}
\label{sec:results}

The following presents the post-fix expected-sensitivity results and
supporting robustness checks. The study is an expected-sensitivity
calculation: no unblinding of a signal region against a background
prediction was performed, and all significances reported below are projections
computed using simulated signal together with a background density model
trained on held-out HTMHT data.

The background model is a conditional flow-matching continuous normalizing
flow trained on the CMS Run~2015D HTMHT MINIAOD open-data sample
\cite{cms_htmht_2015d}. After the hadronic mono-$Z$ preselection the local
extraction selects 1,439,523 events from 20,679,437 raw events in 444
MINIAOD files (selection efficiency 6.96\%). The integrated luminosity used
for normalization is 2.256382381~\invfb, taken from the validated Run2015D
recorded luminosity provenance \cite{cms_data_policy,cms_lumi_guide}.

\begin{table}[H]
  \centering
  \caption{Summary of the processed MINIAOD dataset used for extraction.}
  \begin{tabular}{@{}L{0.62\columnwidth}r@{}}
    \toprule
    Item & Value \\
    \midrule
    Processed MINIAOD files & 444 \\
    Total raw events & 20,679,437 \\
    Selected events (post-selection) & 1,439,523 \\
    Selection efficiency & 6.96\% \\
    \bottomrule
  \end{tabular}
  \label{tab:data-summary}
\end{table}

Full per-file counts (entries and selected events) are provided in
Appendix~\ref{app:filecounts} and summarized above.

Significances are computed with the Asimov counting formula
(Eq.~\eqref{eq:asimov}) described in Section~\ref{sec:sens-methodology},
where $S$ and $B$ denote the weighted signal and background yields above a
threshold on the negative-log-likelihood score. To avoid reporting
optimised cut points driven by very small background populations, the
reported working point is chosen as the best cut satisfying $B\geq20$; the
unconstrained optimum (with $B\ll20$) is retained for diagnostic comparison
but is not used as the headline result. Background yields are estimated from
the held-out 20\% validation split and reweighted to the full selected HTMHT
population to avoid in-sample scoring bias, and the offline HT/MET trigger
proxy described in \S\ref{sec:trigger-proxy} is applied to simulated signal
prior to scoring.

\begin{table*}[t]
  \centering
  \caption{Primary post-fix expected sensitivity for the 32-feature
  flow-matching model. The signal trigger proxy is applied before scoring; the
  reported working point requires $B\geq20$. The final column is the relative
  Poisson uncertainty on the background yield, $1/\sqrt{B}$.}
  \label{tab:primary-results}
  \begin{tabular}{L{3.1cm}L{3.0cm}rrrrrr}
    \toprule
    Signal point & Mediator benchmark & $S$ & $B$ & Best cut & Signal eff. & $Z_A$ & $1/\sqrt{B}$ \\
    \midrule
    axial\_mx10\_mv20 & axial, $m_\chi=10~\GeV$, $m_\mathrm{med}=20~\GeV$
      & 15.76 & 25.00 & 77.05 & 0.250 & 2.89 & 0.20 \\
    axial\_mx50\_mv200 & axial, $m_\chi=50~\GeV$, $m_\mathrm{med}=200~\GeV$
      & 46.90 & 25.00 & 67.43 & 0.203 & 7.62 & 0.20 \\
    vector\_mx1\_mv500 & vector, $m_\chi=1~\GeV$, $m_\mathrm{med}=500~\GeV$
      & 45.38 & 25.00 & 64.97 & 0.218 & 7.41 & 0.20 \\
    \bottomrule
  \end{tabular}
\end{table*}

Table~\ref{tab:primary-results} reports the post-fix baseline numbers. The
light axial benchmark ($m_\chi=10~\GeV$, $m_{\mathrm{med}}=20~\GeV$) yields
an expected significance of $2.89\sigma$, while the heavier axial
($m_\chi=50~\GeV$, $m_{\mathrm{med}}=200~\GeV$) and the vector
($m_\chi=1~\GeV$, $m_{\mathrm{med}}=500~\GeV$) benchmarks reach
$7.62\sigma$ and $7.41\sigma$, respectively. The two heavier benchmarks
therefore provide a clear projected discovery-level sensitivity in this
channel, whereas the light axial point is only moderately sensitive. In all
three cases the reported working point sits at $B\simeq25$ weighted
background events, corresponding to a relative Poisson background
uncertainty of $1/\sqrt{B}\simeq0.20$; the signal efficiency at the working
point is $0.20$--$0.25$, indicating that the selected region retains a
substantial fraction of the signal while keeping the background yield at the
$B\geq20$ floor.

The same analysis also records the unconstrained optima (which occur at very
small background yields) for comparison; these diagnostic values are
significantly larger (e.g. $Z_A\approx4.96,\,10.81,\,10.70$ for the three
benchmarks) but are not used as the principal results due to their
instability under validation-sample fluctuations.

\subsection{Comparison with the leptonic NSF predecessor}

It is instructive to compare the present hadronic flow-matching analysis
with the leptonic NSF study on the same Run~2015D open-data era
\cite{rasineni_nsf_monoz}. The two analyses differ in three principal
respects: the $Z$ decay channel, the density-modelling architecture, and the
statistical interpretation.

First, the leptonic analysis selected $Z\to\ell^{+}\ell^{-}$ events with a
dilepton-mass window $60<m_{\ell\ell}<120~\GeV$, a signal region defined by
$\MET\geq50~\GeV$, $|\Delta\phi(\MET,Z)|>2.5$, and $n_{\mathrm{jets}}\leq1$,
and trained the SM background flows on a control region with
$\MET<50~\GeV$. The present hadronic analysis instead reconstructs the $Z$
from a dijet pair with $70<m_{jj}<110~\GeV$ and $\Delta R_{jj}<2.0$, and
trains the background density directly on the full selected HTMHT
population. The hadronic channel therefore exploits the larger $Z\to
q\bar{q}$ branching fraction at the cost of a substantially more complex
background, which is precisely the regime in which a flexible learned
density model is most valuable.

Second, the density models differ fundamentally. The leptonic analysis used
closed-form NSFs with analytic densities and a two-hypothesis
log-likelihood-ratio score, whereas the present work uses a continuous
flow-matching CNF with an ODE-integrated density and a one-sided
background-NLL discriminant. The CNF's continuous-time formulation avoids
the architectural constraints of invertible coupling layers and is trained
with a simulation-free regression objective, but requires numerical ODE
integration and a stochastic divergence-trace estimate at inference
\cite{lipman_flow_matching,chen_neural_ode,grathwohl_ffjord}.

Third, the statistical treatment differs. The leptonic analysis performed a
simultaneous signal-region/validation-region profile-likelihood fit with
per-channel normalisation nuisances and reported observed and expected
95\% CL upper limits on the signal strength. That fit was dominated by a
residual background-modelling discrepancy in the high-\MET{} tail
($\MET\geq100~\GeV$), which inflated the observed-to-expected limit ratio
by a factor of roughly 7--12 and prevented a clean interpretation of the
fitted signal strength \cite{rasineni_nsf_monoz}. The present analysis
instead reports projected Asimov expected significances
\cite{cowan_asymptotic} computed from a held-out background score
distribution, with a $B\geq20$ working-point floor to avoid the
look-elsewhere bias of an unconstrained discriminant scan. This choice
reflects the projected-sensitivity nature of the present study and avoids
the tail-modelling residual that limited the leptonic interpretation.

\begin{figure*}[t]
  \centering
  \includegraphics[width=0.48\textwidth]{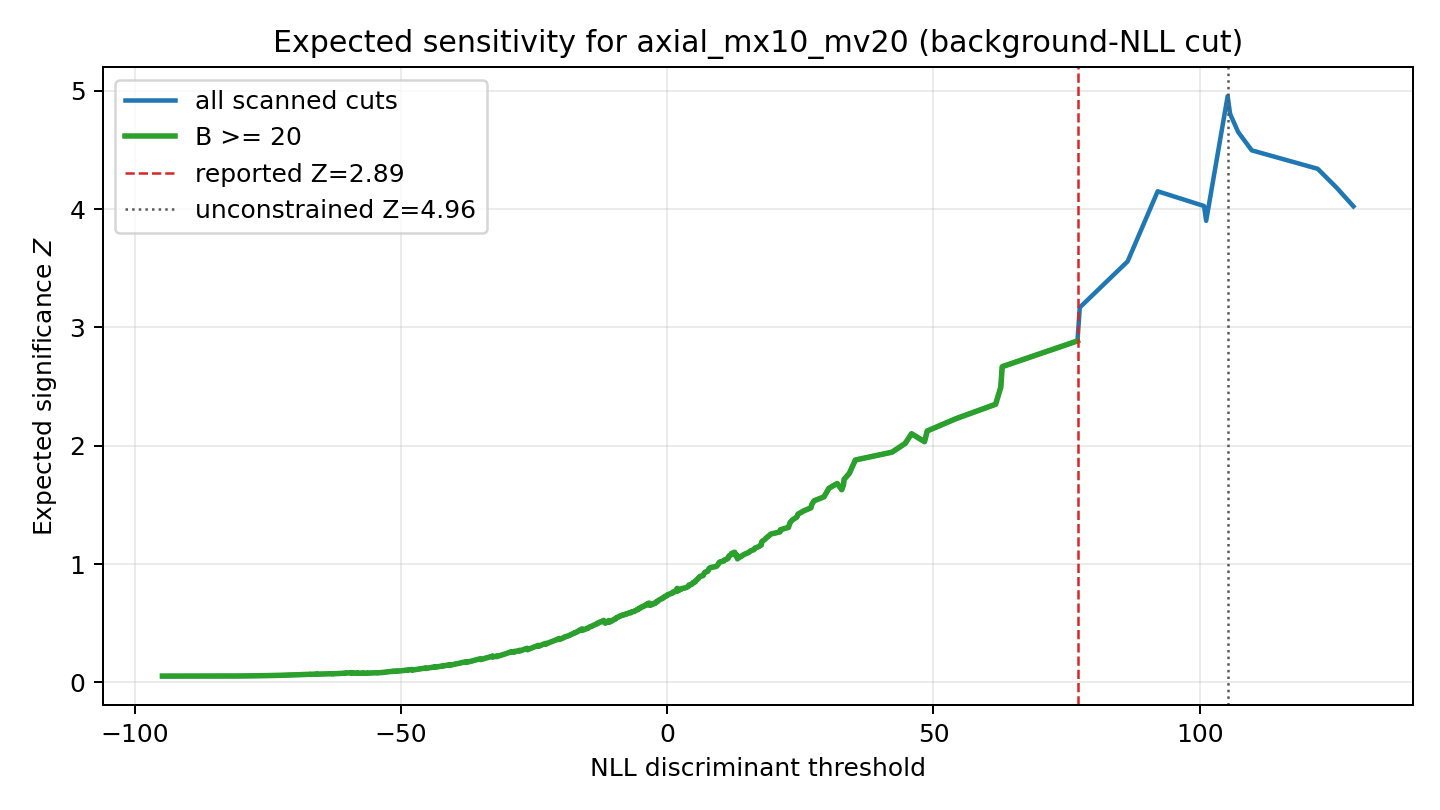}\hfill
  \includegraphics[width=0.48\textwidth]{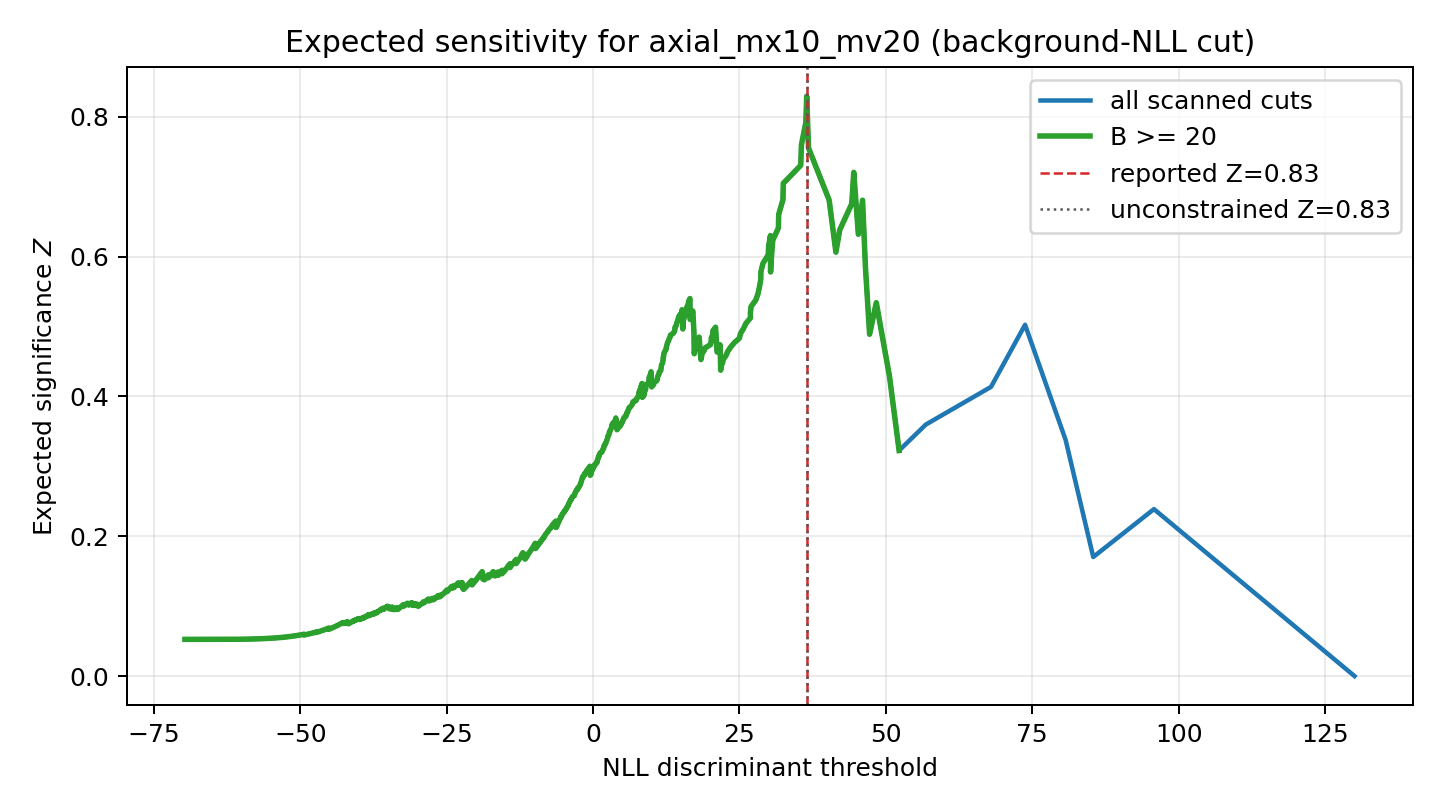}\\[6pt]
  \includegraphics[width=0.48\textwidth]{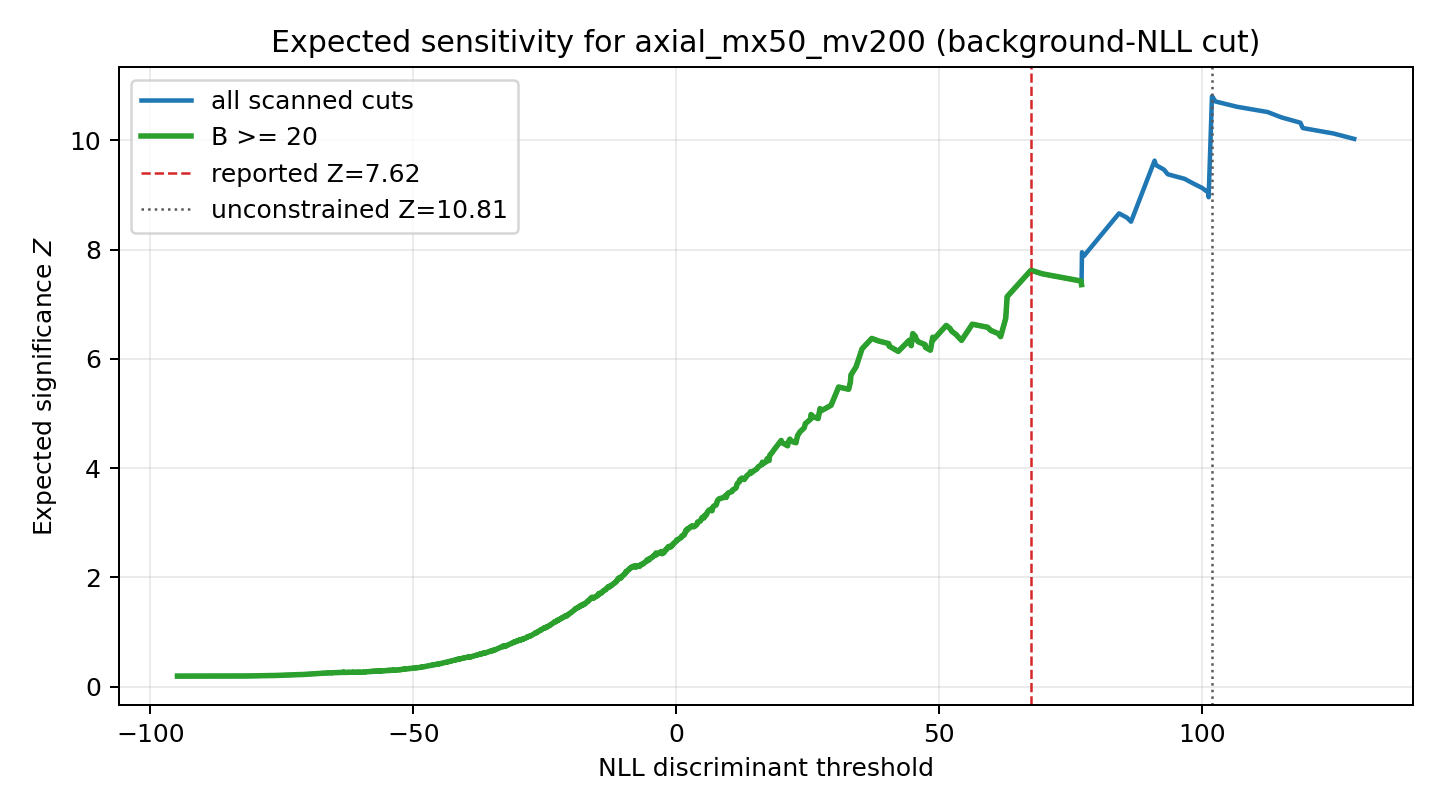}\hfill
  \includegraphics[width=0.48\textwidth]{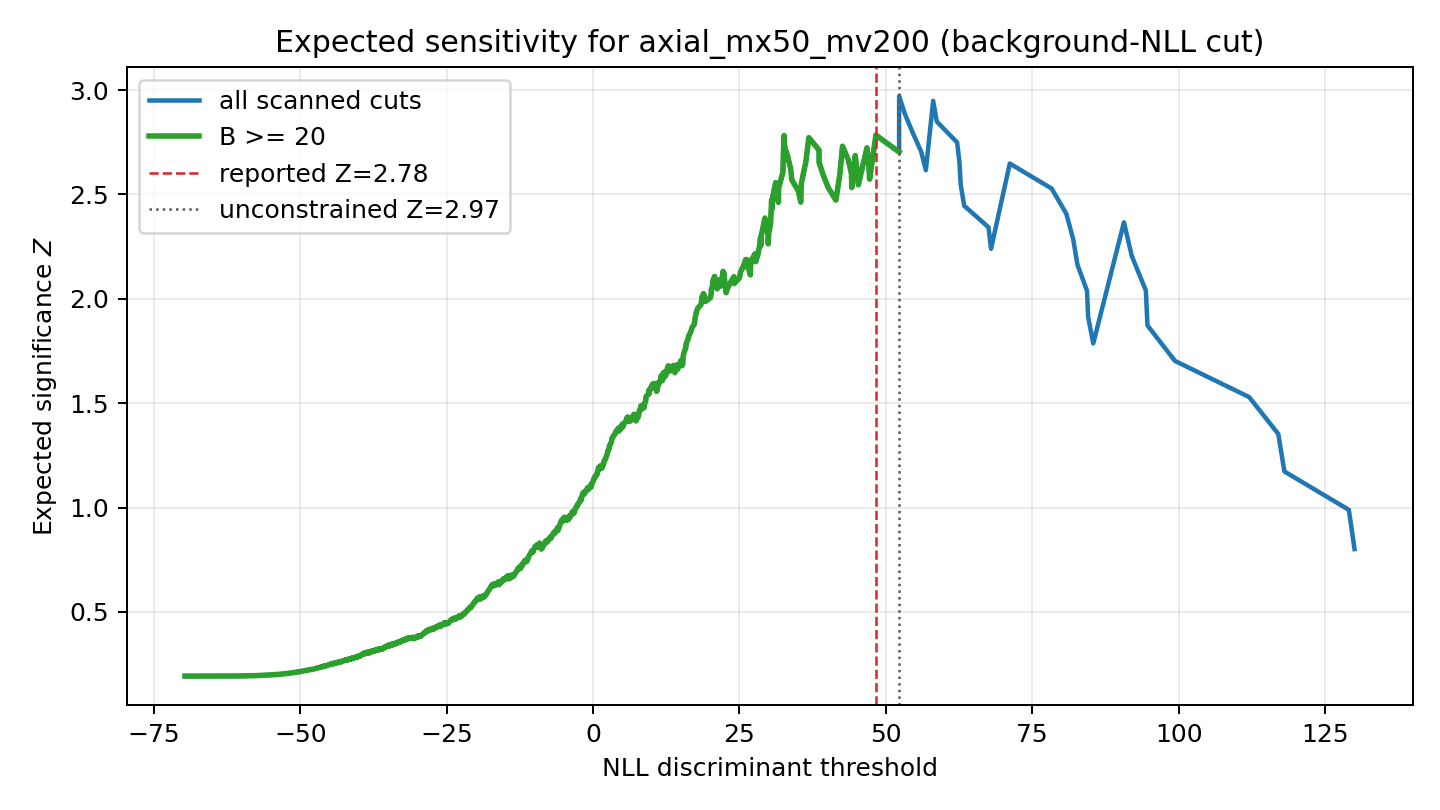}\\[6pt]
  \includegraphics[width=0.48\textwidth]{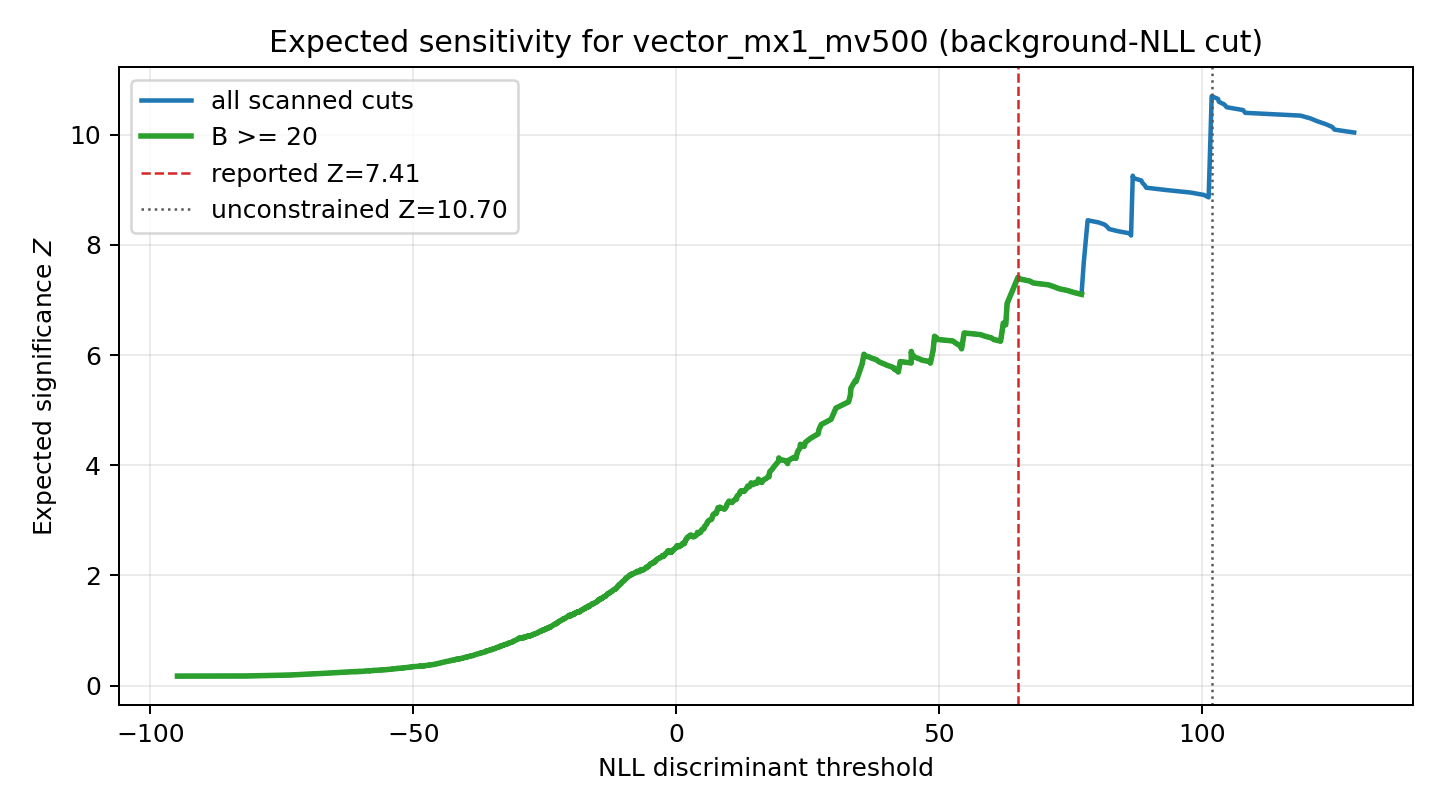}\hfill
  \includegraphics[width=0.48\textwidth]{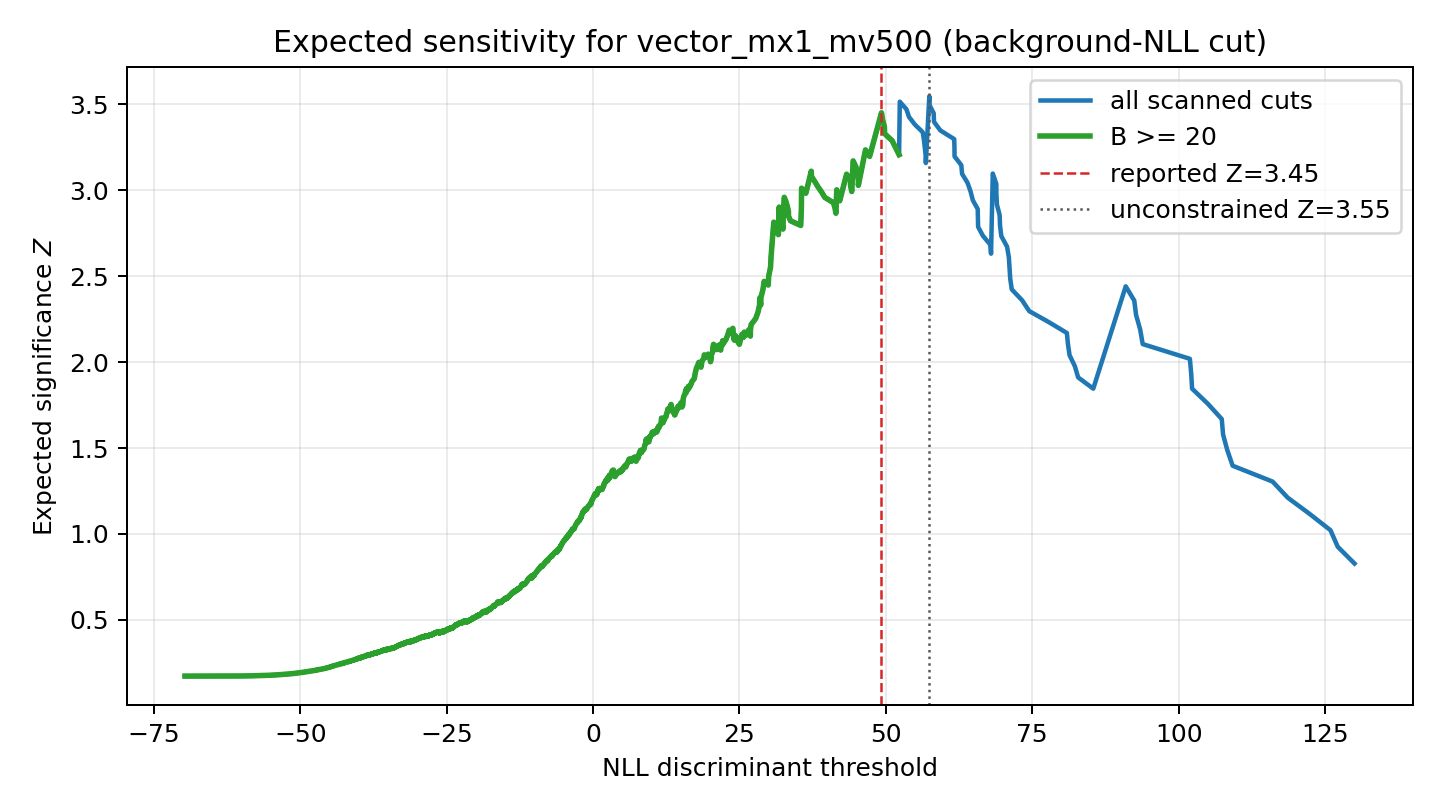}
  \caption{Expected-significance scans versus the NLL discriminant threshold
  for the three benchmark points. Each row pairs the baseline (32-feature)
  model (left) with the extra-jet-drop ablation (26-feature) model (right)
  for one signal point: axial $m_\chi=10/m_\mathrm{med}=20~\GeV$ (top row),
  axial $m_\chi=50/m_\mathrm{med}=200~\GeV$ (middle row), and vector
  $m_\chi=1/m_\mathrm{med}=500~\GeV$ (bottom row). The green curve marks the
  $B\geq20$ constrained scan used for the reported working point, the red
  dashed line the reported optimum, and the grey dotted line the unconstrained
  optimum.}
  \label{fig:sensitivity-scans}
\end{figure*}

\section{Robustness: extra-jet feature ablation}

To assess the contribution of extra-jet kinematics to discrimination, an
ablation study was performed in which six padded extra-jet columns were
removed from the training feature set: the leading and subleading extra-jet
$p_T$ and $\eta$ (\texttt{jet1\_pt}, \texttt{jet1\_eta},
\texttt{jet2\_pt}, \texttt{jet2\_eta}) and the two corresponding
$\Delta\phi(\MET,\mathrm{jet})$ variables
(\texttt{dphi\_met\_jet1}, \texttt{dphi\_met\_jet2}). This reduces the
feature dimension from $D=32$ to $D=26$. The ablation model was trained from
scratch with the same architecture, hyperparameters, and held-out scoring
procedure as the baseline; its best checkpoint is at epoch 165 (validation
NLL $\approx -43.63$), and the closure validation
(Section~\ref{sec:validation}) confirms that the ablation model is not
undertrained---every retained feature closes as well as in the baseline.

\begin{table*}[t]
  \centering
  \caption{Effect of dropping the extra-jet kinematic features. The ablation
  working point is selected with the same $B\geq20$ floor as the baseline.}
  \label{tab:ablation-results}
  \begin{tabular}{L{2.6cm}rrrrrrr}
    \toprule
    & \multicolumn{3}{c}{Baseline ($D=32$)} & \multicolumn{3}{c}{Ablation ($D=26$)} & \\
    \cmidrule(lr){2-4}\cmidrule(lr){5-7}
    Signal point & $S$ & $B$ & $Z_A$ & $S$ & $B$ & $Z_A$ & Rel.\ drop \\
    \midrule
    axial 10/20  & 15.76 & 25.00 & 2.89 & 5.98  & 50.00 & 0.83 & 71.3\% \\
    axial 50/200 & 46.90 & 25.00 & 7.62 & 15.16 & 25.00 & 2.78 & 63.5\% \\
    vector 1/500 & 45.38 & 25.00 & 7.41 & 19.13 & 25.00 & 3.45 & 53.5\% \\
    \bottomrule
  \end{tabular}
\end{table*}

\begin{figure*}[t]
  \centering
  \includegraphics[width=0.48\textwidth]{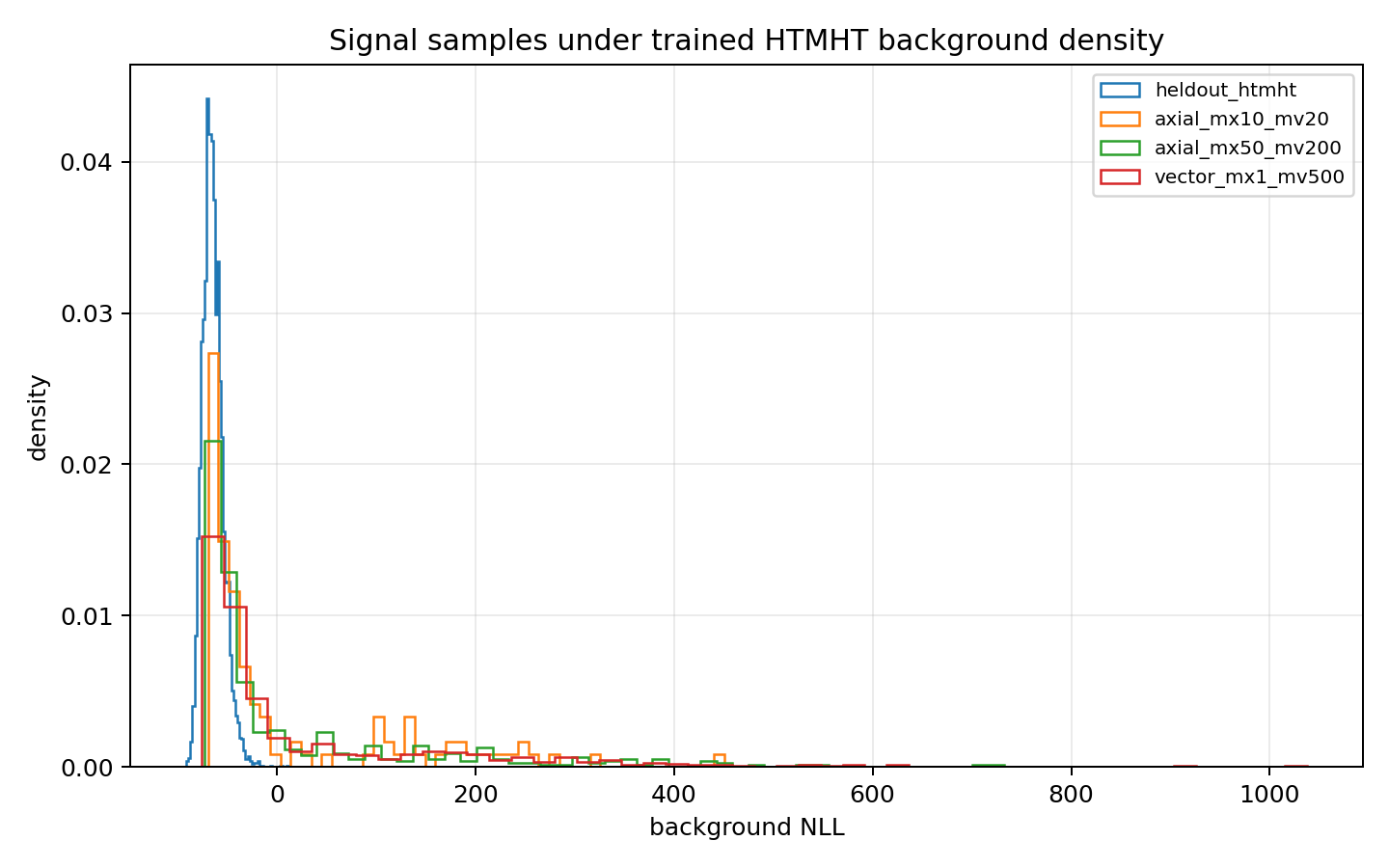}\hfill
  \includegraphics[width=0.48\textwidth]{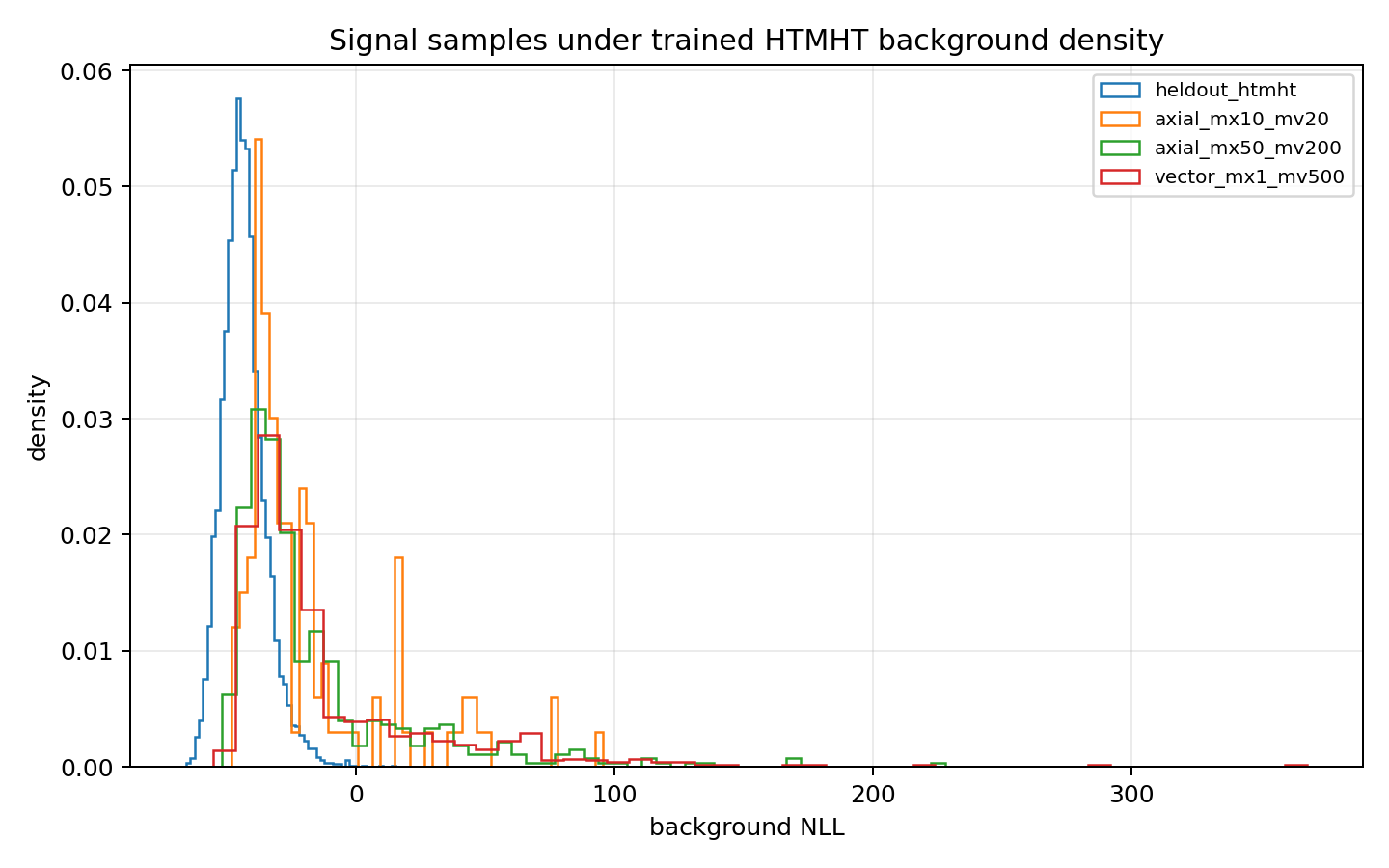}
  \caption{Signal-versus-background NLL distributions for the baseline
  (left, $D=32$) and extra-jet-drop ablation (right, $D=26$) models. In the
  ablation model the \texttt{axial\_mx10\_mv20} signal distribution (orange)
  overlaps almost entirely with the background peak (blue), explaining its
  collapse to sub-$1\sigma$ significance.}
  \label{fig:signal-vs-background-nll}
\end{figure*}

Table~\ref{tab:ablation-results} shows that removing detailed extra-jet
information reduces the expected significance at all three benchmark points,
with relative drops of 53--71\%. The largest impact is on the light axial
benchmark ($m_\chi=10~\GeV$, $m_{\mathrm{med}}=20~\GeV$), whose significance
collapses from $2.89\sigma$ to $0.83\sigma$; in the ablation model the
$B\geq20$ floor forces the working point to a higher background yield
($B\simeq50$) at a lower signal efficiency, indicating that the signal NLL
distribution has lost separation from the background. This is visible in
the signal-versus-background NLL overlap plot
(Figure~\ref{fig:signal-vs-background-nll}): in the ablation model the
\texttt{axial\_mx10\_mv20} signal NLL distribution sits almost entirely on
top of the background peak, barely distinguishable, whereas the other two
benchmarks retain a visible (if reduced) tail past the background peak.

This behavior supports the interpretation that the extra-jet kinematics
carry genuine discriminating information for the hadronic mono-$Z$ topology,
plausibly arising from initial-state-radiation jet topology differences
between mono-$Z$ dark-matter production and generic multijet HTMHT
background, rather than being an artifact of the (now-fixed) sentinel
imputation. The count feature \texttt{n\_extra\_jets} alone is
insufficient to recover this discriminating power; the detailed per-jet
kinematics matter.

\section{Limitations}
\label{sec:limitations}

Several caveats should be emphasized when interpreting these results.

\textit{Projected sensitivity, not observation.}
The study reports projected, expected significances rather than observed
excesses. The background density is trained on real selected HTMHT events,
but $S$, $B$, and $Z_A$ are obtained by combining the held-out background
score distribution with simulated signal. No unblinding of a signal region
against a background prediction was performed at any point.

\textit{Trigger proxy.}
The signal trigger treatment is an offline proxy implemented to approximate
the dominant HT/MET/monojet HLT logic in the recorded dataset. Delphes does
not emulate the full HLT turn-on, and a complete experimental
reinterpretation would require trigger-efficiency parameterization and
detector systematic uncertainties \cite{delphes3}. The proxy efficiency is
100\% for all three benchmarks after offline selection, which is likely an
overestimate of the true trigger acceptance for the lighter benchmarks.

\textit{No background systematic uncertainties.}
Only the statistical (Poisson) background uncertainty, $1/\sqrt{B}$, is
propagated. No systematic uncertainty on the background density model---for
example from the choice of training features, the ODE solver tolerance, or
the finite validation sample---is included. This is an expected-sensitivity
study, not a full profile-likelihood limit-setting analysis, and the
reported $Z_A$ values should be interpreted accordingly.

\textit{Signal simulation gaps.}
The Delphes signal simulation does not include a pileup overlay, so the
\texttt{n\_vertices} and \texttt{rho} features are zero placeholders for
signal and are excluded from training. The b-tag discriminator semantics
also differ between Delphes (binary flag) and data (continuous CSVv2 score),
so the b-tag-related columns are excluded from the feature set. A full
treatment would require pileup overlay and a Delphes card with a compatible
b-tag discriminator output.

\textit{Delphes fast-simulation scope.}
The signal samples rely on the Delphes~3 fast simulation
\cite{delphes3}, which is designed for phenomenological studies rather than
for reproducing the full detector response of a specific experiment. As
documented in the Delphes~3 paper, the framework parameterises the detector
response with smearing functions and a simplified particle-flow
reconstruction, and it does not simulate fake rates for leptons and photons
or a full trigger emulation \cite{delphes3}. The Delphes validation against
CMS and ATLAS shows good agreement for jet and \MET{} resolutions at
$p_T > 30~\GeV$ \cite{delphes3}, which is the regime relevant to the
hadronic mono-$Z$ selection used here. Nevertheless, the absence of a
pileup overlay in the CMS card and the parametric b-tagging are inherent
limitations of the fast-simulation approach, and the corresponding features
are therefore excluded from the density model rather than treated as
mismodelled inputs. This is consistent with the Delphes authors' own
statement that the framework is not meant for advanced detector studies,
for which more accurate tools are required \cite{delphes3}.

\textit{Incomplete signal cutflow.}
Per-stage signal-side cutflow counters (trigger proxy, jet kinematics,
$m_{jj}$ window, $\Delta R$, b-veto) were added to the signal-generation
notebook but the resulting per-stage CSVs have not been reviewed to diagnose
why the \texttt{axial\_mx10\_mv20} benchmark has a lower raw selection
efficiency than the other two points. This diagnosis remains open.

\textit{Continuous-flow structural limitation.}
Continuous normalizing flows cannot represent exact discrete-plus-continuous
mixtures (for example, the point mass at zero for padded extra-jet $p_T$
combined with a continuous distribution for present jets). The
sentinel-imputation fix applied to undefined angular features removes the
largest imputation artifact, but the $p_T$ point-mass limitation remains a
structural caveat of the continuous-density approach and would require a
discrete-plus-continuous hybrid architecture to resolve fully.

\section{Conclusion}

We have presented a projected sensitivity study for hadronic mono-$Z$ dark
matter production using CMS Run~2015D HTMHT open data corresponding to
$2.256382381~\invfb$. A conditional flow-matching continuous normalizing
flow is trained on 1{,}439{,}523 selected background events to model the
hadronic mono-$Z$ background density, and per-event negative-log-likelihood
scores serve as the discriminant between background and simulated
dark-matter signal. The methodology includes sentinel imputation for
undefined extra-jet angular features, held-out-only background scoring with
population reweighting, a signal-side offline trigger proxy, and a
minimum-background-yield ($B\geq20$) working-point floor to avoid
look-elsewhere bias in the NLL-cut scan.

Under this procedure the post-fix baseline analysis yields expected
significances of $2.89\sigma$, $7.62\sigma$, and $7.41\sigma$ for the three
retained simplified-model benchmarks (axial $m_\chi=10/m_{\mathrm{med}}=20$,
axial $m_\chi=50/m_{\mathrm{med}}=200$, and vector
$m_\chi=1/m_{\mathrm{med}}=500~\GeV$). An ablation study that removes the
six detailed extra-jet kinematic features reduces the expected significance
by 53--71\%, demonstrating that extra-jet topology---plausibly from
initial-state radiation---carries substantial genuine discriminating power
in the hadronic mono-$Z$ channel beyond what the jet multiplicity alone
provides.

This work also extends our earlier leptonic mono-$Z$ study
\cite{rasineni_nsf_monoz} in three ways. Methodologically, it replaces the
closed-form neural spline flow with a continuous flow-matching normalizing
flow, trading exact analytic densities for a more flexible continuous-time
model trained with a simulation-free objective. Statistically, it replaces
the two-hypothesis likelihood-ratio score and profile-likelihood fit with a
single background-density discriminant evaluated on a held-out validation
split and converted into an Asimov expected significance, avoiding the
high-\MET{} tail-modelling residual that dominated the leptonic
interpretation. Physically, it moves from the clean leptonic $Z$ decay to
the higher-rate hadronic channel, where the substantially more complex
multijet background makes a flexible learned density model particularly
valuable and where the extra-jet topology provides a genuine discriminating
handle.

Future work should extend this study to a full CLs/profile-likelihood
treatment with background systematic uncertainties, replace the offline
trigger proxy with a data-driven HLT efficiency measurement, add pileup
overlay to the signal simulation, diagnose the per-stage signal cutflow for
the light axial benchmark, and perform a direct head-to-head comparison of
the flow-matching and neural-spline-flow density models on the same hadronic
channel and dataset \cite{rasineni_nsf_monoz}.

\section*{Declarations}
\addcontentsline{toc}{section}{Declarations}

\noindent\textbf{Funding:} The authors declare that no funding was received for this work.

\noindent\textbf{Ethics, Consent to Participate, and Consent to Publish declarations:} Not applicable.

\noindent\textbf{Conflict of interest:} The authors declare no competing interests.

\noindent\textbf{Data availability:} The datasets analysed are publicly available CMS Open Data.

\clearpage
\onecolumn
\raggedbottom
\appendix
\begin{center}
\Large\textbf{Appendix}
\end{center}
\vspace{1em}

\section{HLT paths and trigger details}
\label{app:hlt}

The primary HLT paths used to select the background sample (and used to
construct the signal-side offline trigger proxy) are listed here for
completeness:

\begin{itemize}
  \item \texttt{HLT\_PFHT350\_v}
  \item \texttt{HLT\_PFHT475\_v}
  \item \texttt{HLT\_PFMETNoMu90\_PFMHTNoMu90\_IDTight\_v}
  \item \texttt{HLT\_PFMETNoMu120\_PFMHTNoMu120\_IDTight\_v}
  \item \texttt{HLT\_MonoCentralPFJet80\_PFMETNoMu90\_PFMHTNoMu90\_IDTight\_v}
\end{itemize}

In the 2015 MiniAOD files the \texttt{selectedPatTrigger.pathNames\_} branch
is often empty, so the trigger decision is reconstructed from the
\texttt{filterLabels\_} branch using the following HLT filter tokens:
\texttt{hltPFHT}, \texttt{hltPFMET}, \texttt{hltPFMHT}, and
\texttt{hltMonoCentralPFJet}. The trigger audit confirms that
18{,}875{,}596 of 20{,}679{,}437 raw events (91.3\%) pass at least one of
the five HLT paths, all identified via the filter-label fallback.

The signal offline trigger proxy applied to Delphes signal samples is
$H_T>350$~\GeV OR $\MET>120$~\GeV OR ($\MET>90$~\GeV and leading jet
$p_T>80$~\GeV), as defined in Section~\ref{sec:trigger-proxy}.

The b-tag veto uses the CSVv2 medium working point at a discriminator value
of 0.800, corresponding to the 2015/76X working point for the
\texttt{pfCombinedInclusiveSecondaryVertexV2BJetTags} discriminator. The
veto is applied only to extra jets beyond the two $Z$-candidate jets; the
$Z$-daughter jets are not subject to the b-tag requirement.

\section{Feature definitions and extraction formulas}
\label{app:features}

This appendix tabulates the 41-branch data/signal feature schema and the
formulas used to compute each branch. Notation: $\vec p = (p_x,p_y,p_z)$,
$p_T = \sqrt{p_x^2+p_y^2}$, and $\Delta\phi(a,b)$ denotes the azimuthal
separation wrapped to $(-\pi,\pi]$.

The offline selection parameters are: leading and subleading $Z$-candidate
jet $p_T > 30~\GeV$, $|\eta| < 2.4$, $70 < m_{jj} < 110~\GeV$,
$\Delta R_{jj} < 2.0$, and a b-tag veto on extra jets at CSVv2 medium
(0.800). The 41-branch data/signal schema is reduced to 32 training
features by excluding nine columns: \texttt{source\_file\_idx},
\texttt{source\_entry} (audit bookkeeping), \texttt{trigger\_pass},
\texttt{btag\_pass}, \texttt{n\_bjets} (trigger/veto flags),
\texttt{max\_btag\_csvv2}, \texttt{btag\_wp\_medium} (b-tag schema
mismatch), and \texttt{n\_vertices}, \texttt{rho} (pileup mismatch).

\small
\renewcommand{\arraystretch}{1.15}
\setlength{\LTleft}{0pt}
\setlength{\LTright}{0pt}
\setlength{\LTpre}{0pt plus 1pt minus 0pt}
\setlength{\LTpost}{0pt plus 1pt minus 0pt}
\begin{longtable}{@{}L{0.28\textwidth}L{0.67\textwidth}@{}}
\caption{Feature schema (41 branches) and extraction formulas. Nine columns excluded from training are annotated ``(not trained)''.}\label{tab:feature-formulas}\\
\toprule
Feature & Formula \\
\midrule
\endfirsthead
\toprule
Feature & Formula \\
\midrule
\endhead
\midrule
\multicolumn{2}{r}{Continued on next page}\\
\endfoot
\bottomrule
\endlastfoot
\texttt{met\_pt} & $\MET = \sqrt{p_{x,\text{miss}}^2 + p_{y,\text{miss}}^2}$ \\
\texttt{met\_phi} & $\mathrm{atan2}(p_{y,\text{miss}},p_{x,\text{miss}})$ \\
\texttt{min\_dphi\_met\_jets} & $\min_i\,|\Delta\phi(\mathrm{MET},\mathrm{jet}_i)|$ \\
\texttt{dphi\_met\_jet1} & $\Delta\phi(\mathrm{MET},\mathrm{jet}_1)$ \\
\texttt{dphi\_met\_jet2} & $\Delta\phi(\mathrm{MET},\mathrm{jet}_2)$ \\
\texttt{met\_over\_sqrtHT} & $\MET/\sqrt{HT}$ \\
\texttt{hadronic\_recoil\_pt} & $\sqrt{\text{recoil}_x^2 + \text{recoil}_y^2}$ \\
\texttt{m\_jj} & $\sqrt{(E_1+E_2)^2 - |\vec p_1 + \vec p_2|^2}$ \\
\texttt{pt\_Z} & $p_{T,Z}=\sqrt{p_{x,Z}^2+p_{y,Z}^2}$ with $p_Z=p_1+p_2$ \\
\texttt{eta\_Z\_pseudo} & $\eta_Z=-\ln\tan(\theta_Z/2)$, $\theta_Z=\arctan2(p_{T,Z},p_{z,Z})$ \\
\texttt{phi\_Z} & $\operatorname{atan2}(p_{y,Z},p_{x,Z})$ \\
\texttt{dphi\_met\_Z} & $\Delta\phi(\mathrm{MET},Z)$ \\
\texttt{met\_over\_ptZ} & $\MET/p_{T,Z}$ \\
\texttt{abs\_met\_minus\_ptZ} & $|\MET-p_{T,Z}|$ \\
\texttt{abs\_mjj\_minus\_mZ} & $|m_{jj}-m_Z|$, with $m_Z=91.1876~\GeV$ \\
\texttt{u\_parallel} & $\text{recoil}_x z_{\mathrm{unit},x}+\text{recoil}_y z_{\mathrm{unit},y}$ \\
\texttt{u\_perp} & $-\text{recoil}_x z_{\mathrm{unit},y}+\text{recoil}_y z_{\mathrm{unit},x}$ \\
\texttt{u\_parallel\_over\_ptZ} & $u_{\parallel}/p_{T,Z}$ \\
\texttt{u\_perp\_over\_ptZ} & $u_{\perp}/p_{T,Z}$ \\
\texttt{zjet1\_pt} & $p_{T,\text{leading Z-daughter jet}}$ \\
\texttt{zjet1\_eta} & $\eta_{\text{leading Z-daughter jet}}$ \\
\texttt{zjet2\_pt} & $p_{T,\text{subleading Z-daughter jet}}$ \\
\texttt{zjet2\_eta} & $\eta_{\text{subleading Z-daughter jet}}$ \\
\texttt{deltaR\_zjj} & $\sqrt{(\eta_{j1}-\eta_{j2})^2+\Delta\phi(\phi_{j1},\phi_{j2})^2}$ \\
\texttt{deltaphi\_zjj} & $\Delta\phi(\phi_{j1},\phi_{j2})$ \\
\texttt{n\_jets} & $N_{\text{jets}}$ \\
\texttt{n\_extra\_jets} & $N_{\text{jets}}-2$ \\
\texttt{jet1\_pt} & $p_{T,\text{leading extra jet}}$ \\
\texttt{jet1\_eta} & $\eta_{\text{leading extra jet}}$ (sentinel when absent) \\
\texttt{jet2\_pt} & $p_{T,\text{subleading extra jet}}$ \\
\texttt{jet2\_eta} & $\eta_{\text{subleading extra jet}}$ (sentinel when absent) \\
\texttt{HT} & $HT=\sum_{\text{jets}} p_{T,\text{jet}}$ \\
\texttt{n\_bjets} & $N_{\text{b-jets}}$, count of extra jets with $\mathrm{CSVv2}\geq0.800$; not trained (veto flag) \\
\texttt{n\_vertices} & $N_{\text{vtx}}$, number of reconstructed primary vertices from \texttt{recoVertexs\_offlineSlimmedPrimaryVertices}; not trained (pileup mismatch) \\
\texttt{rho} & $\rho_{\text{FastJet}}=\texttt{double\_fixedGridRhoFastjetAll}$ (pileup energy density); not trained (pileup mismatch) \\
\texttt{source\_file\_idx} & Integer index of source MINIAOD file in traversal order; not trained (audit) \\
\texttt{source\_entry} & Entry index within source MINIAOD file; not trained (audit) \\
\texttt{trigger\_pass} & Binary HLT decision: $\mathbb{1}[\text{event fires}\geq1\text{ path in App.~\ref{app:hlt}}]$ (via \texttt{pathNames\_} or \texttt{filterLabels\_} fallback); not trained (trigger flag) \\
\texttt{btag\_pass} & Binary selection flag: $\mathbb{1}[\text{no extra jet has }\mathrm{CSVv2}\geq0.800]$; not trained (veto flag) \\
\texttt{max\_btag\_csvv2} & $\max_{i\in\text{extra jets}}\mathrm{CSVv2}_i$; $-1.0$ if no extra jets exist; not trained (b-tag schema mismatch) \\
\texttt{btag\_wp\_medium} & Constant $0.800$, the CSVv2 medium working-point threshold (\texttt{BTAG\_MEDIUM\_WP}); not trained (b-tag schema mismatch) \\
\end{longtable}
\normalsize

\clearpage

\section{Per-file event counts and audit}
\label{app:filecounts}

During dataset traversal, per-MINIAOD-file raw and selected event counts
were tracked. For the dataset used in this study 444 MINIAOD files were
processed, yielding 20{,}679{,}437 raw events and 1{,}439{,}523 selected
events after the analysis selection (overall selection efficiency 6.96\%).
No chunks were skipped and no events were lost to processing errors. The
trigger audit records 18{,}875{,}596 events passing at least one HLT path
(91.3\% of raw events), all identified via the filter-label fallback. The
b-tag decoder audit records 204{,}250{,}006 jets decoded across all events,
with 14 discriminator labels observed in the \texttt{pairDiscriVector\_}
payload; the primary discriminator
\texttt{pfCombinedInclusiveSecondaryVertexV2BJetTags} has a score range of
$[-10, 1]$. Per-file raw entries vary from a few hundred to a few tens of
thousands; typical per-file counts are of order $10^{3}$--$10^{4}$. For
brevity the manuscript omits the full per-file listing; complete per-file
counts are available with the analysis code
(Section~\ref{sec:reproducibility}).

\subsection{Data-side cutflow}

Table~\ref{tab:data-cutflow} summarises the data-side event selection
stages. The trigger stage applies the five unprescaled HLT paths of
Appendix~\ref{app:hlt}; the offline stage applies the hadronic mono-$Z$
selection described in Section~\ref{sec:data} (jet $p_T$ and $\eta$
requirements, the $m_{jj}$ window, $\Delta R_{jj}$, and the b-tag veto on
extra jets). The overall selection efficiency is the ratio of the final
selected population to the raw event count.

\begin{table}[H]
  \centering
  \caption{Data-side cutflow for the CMS Run~2015D HTMHT open dataset.}
  \label{tab:data-cutflow}
  \begin{tabular}{@{}L{0.55\columnwidth}rr@{}}
    \toprule
    Stage & Events & Efficiency \\
    \midrule
    Raw events processed & 20{,}679{,}437 & 100\% \\
    Pass $\geq$1 HLT path & 18{,}875{,}596 & 91.3\% \\
    Pass offline hadronic mono-$Z$ selection & 1{,}439{,}523 & 6.96\% \\
    \bottomrule
  \end{tabular}
\end{table}

The signal-side cutflow is not tabulated here because the per-stage
signal-generation counters have not yet been reviewed (see
Section~\ref{sec:limitations}); the final offline post-selection yields
for the three benchmarks are 116, 487, and 848 events, respectively.

\section{Full closure figures}
\label{app:closures}

The full set of per-feature closure plots not shown in the main text is
collected below. Representative closure plots for $m_{jj}$, $\MET$, $H_T$,
and $\Delta R_{Zjj}$ appear in Figure~\ref{fig:closure-representative};
additional kinematic closure panels appear in
Figure~\ref{fig:closure-kinematics}; the sentinel-imputation panels appear
in Figure~\ref{fig:imputation-comparison}; the validation-NLL curves appear
in Figure~\ref{fig:validation-nll}; and the sensitivity and
signal-versus-background NLL figures appear in
Figures~\ref{fig:sensitivity-scans} and~\ref{fig:signal-vs-background-nll}.
The remaining per-feature closure panels are collected below.

\subsection*{Remaining baseline closure figures}

\begin{figure}[p]
  \centering
  \includegraphics[width=0.48\textwidth]{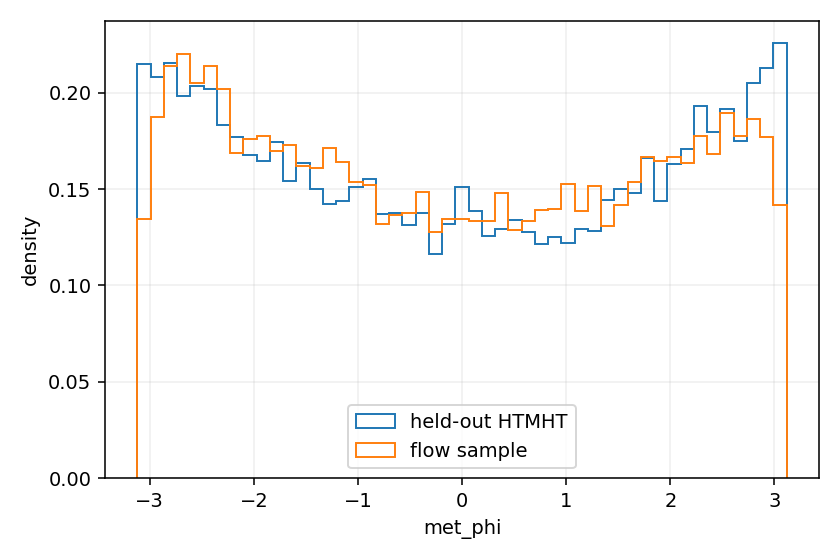}\hfill
  \includegraphics[width=0.48\textwidth]{images/closure_dphi_met_jet1.png}\\[6pt]
  \includegraphics[width=0.48\textwidth]{images/closure_dphi_met_jet2.png}\hfill
  \includegraphics[width=0.48\textwidth]{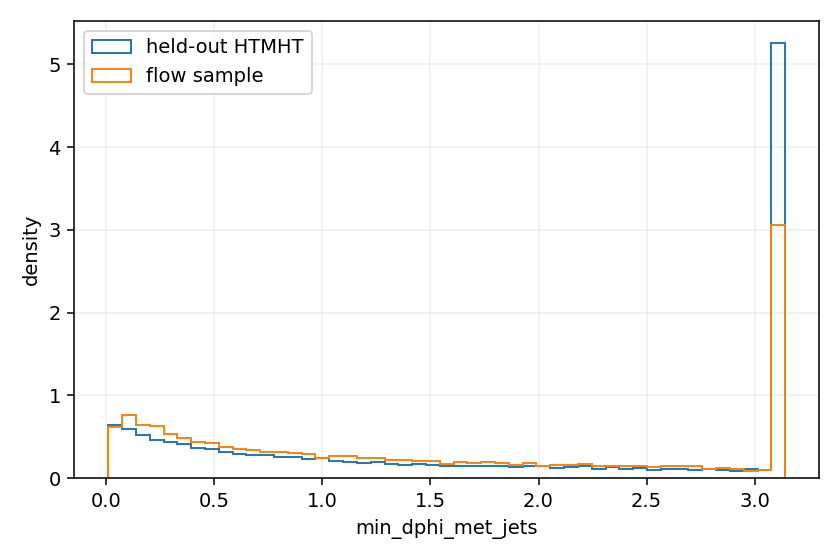}
  \caption{Baseline closure plots for the missing-momentum observables and the undefined extra-jet angular features.}
\end{figure}

\begin{figure}[p]
  \centering
  \includegraphics[width=0.48\textwidth]{images/closure_met_over_sqrtHT.png}\hfill
  \includegraphics[width=0.48\textwidth]{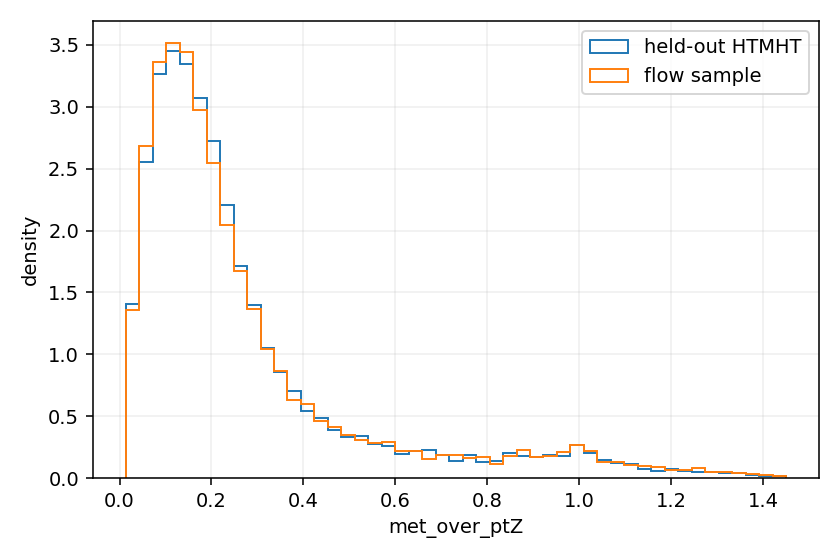}\\[6pt]
  \includegraphics[width=0.48\textwidth]{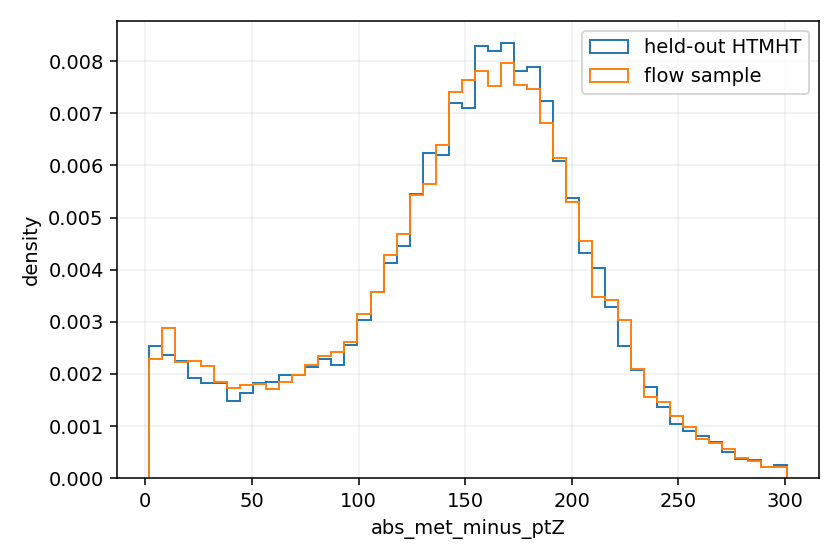}\hfill
  \includegraphics[width=0.48\textwidth]{images/closure_abs_mjj_minus_mZ.png}
  \caption{Baseline closure plots for the MET-to-scale ratios and angular proximity summary.}
\end{figure}

\begin{figure}[p]
  \centering
  \includegraphics[width=0.48\textwidth]{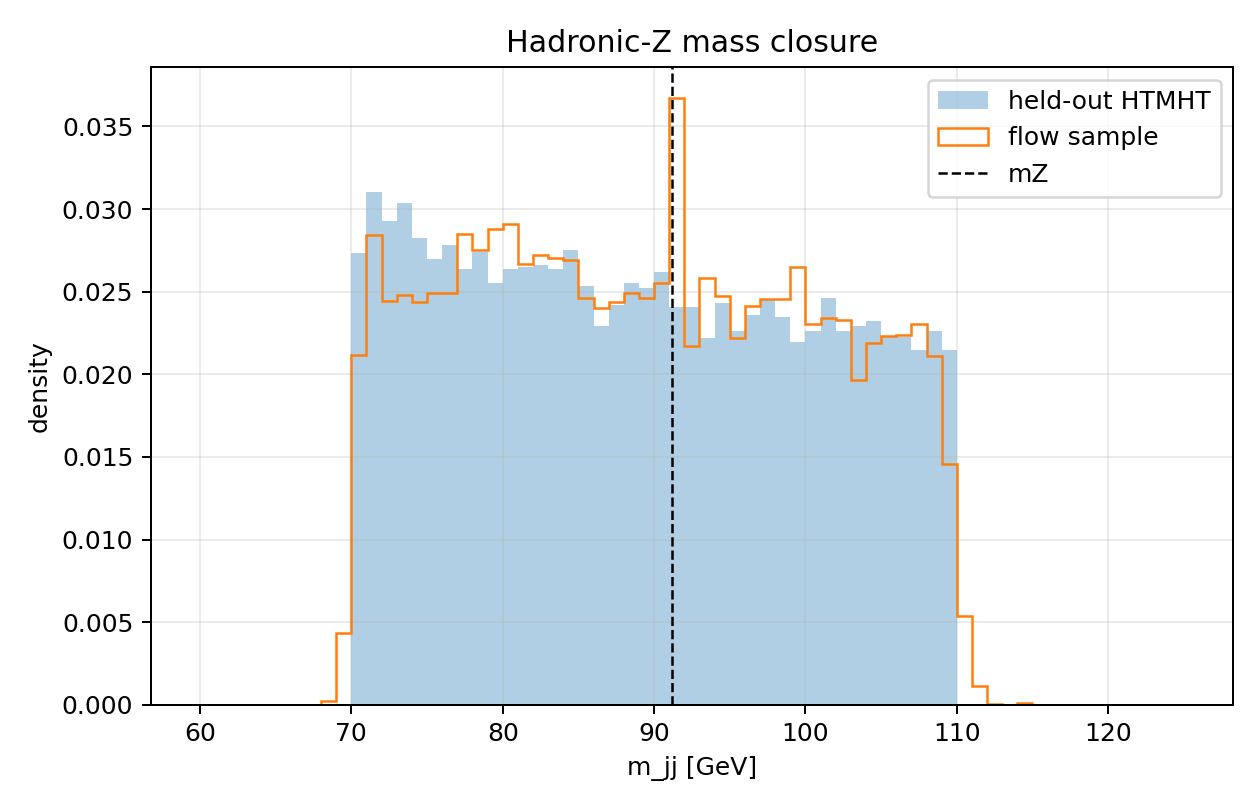}\hfill
  \includegraphics[width=0.48\textwidth]{images/closure_pt_Z.png}\\[6pt]
  \includegraphics[width=0.48\textwidth]{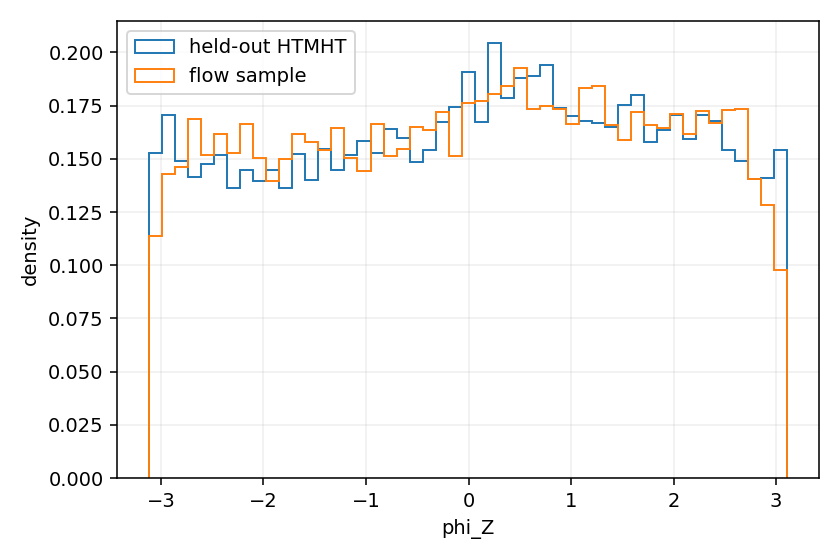}\hfill
  \includegraphics[width=0.48\textwidth]{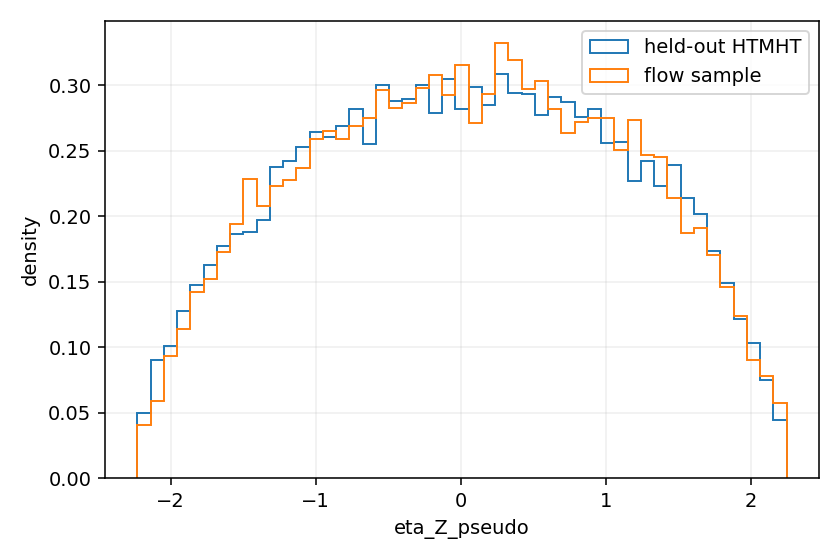}
  \caption{Baseline closure plots for the hadronic-$Z$ mass window observables and reconstructed $Z$ kinematics.}
\end{figure}

\begin{figure}[p]
  \centering
  \includegraphics[width=0.48\textwidth]{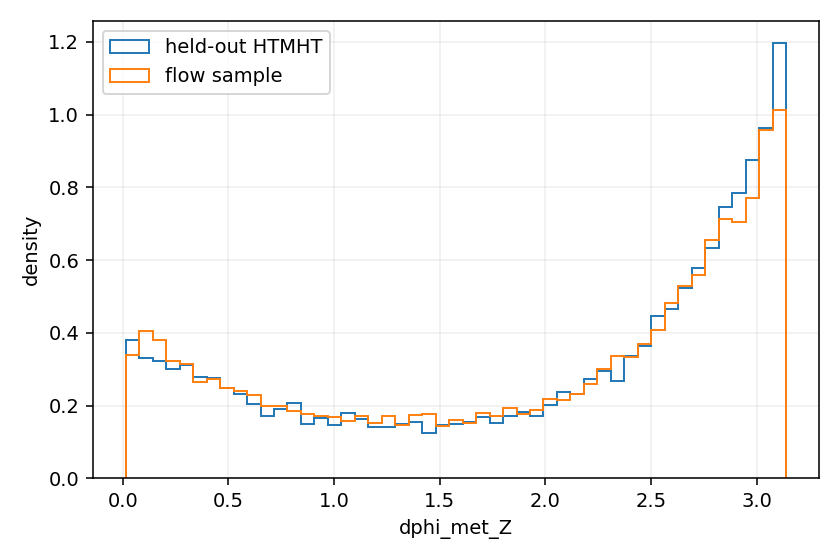}\hfill
  \includegraphics[width=0.48\textwidth]{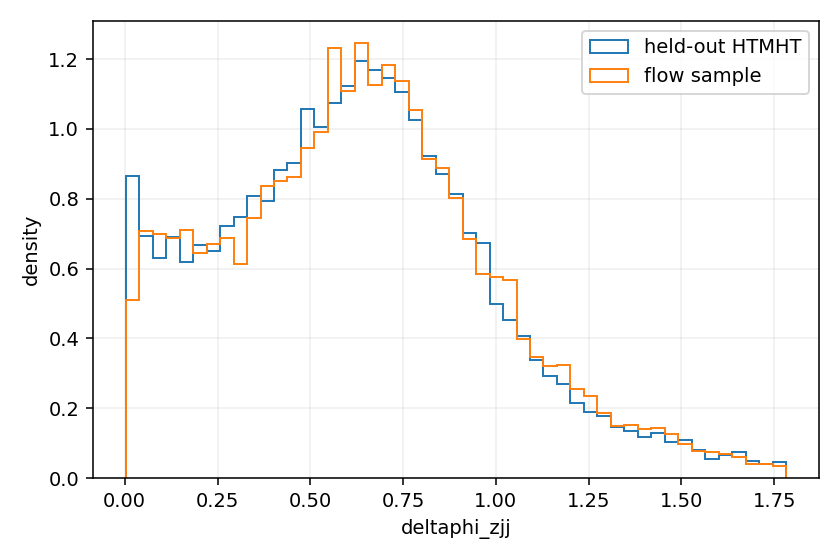}\\[6pt]
  \includegraphics[width=0.48\textwidth]{images/closure_hadronic_recoil_pt.png}\hfill
  \includegraphics[width=0.48\textwidth]{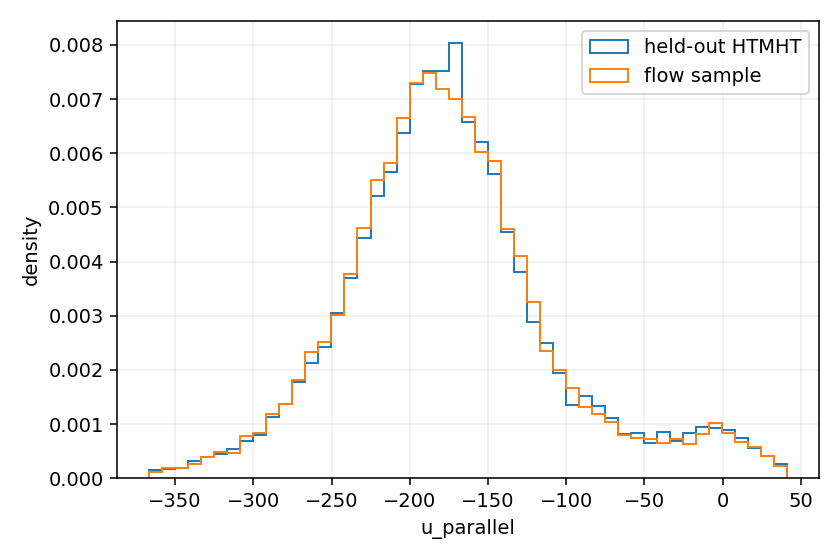}
  \caption{Baseline closure plots for the reconstructed $Z$ direction, dijet geometry, and recoil decomposition.}
\end{figure}

\begin{figure}[p]
  \centering
  \includegraphics[width=0.48\textwidth]{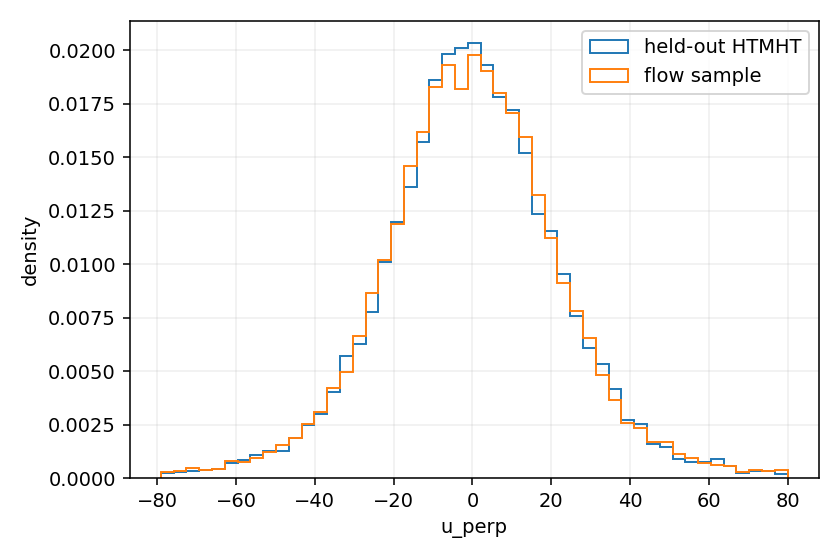}\hfill
  \includegraphics[width=0.48\textwidth]{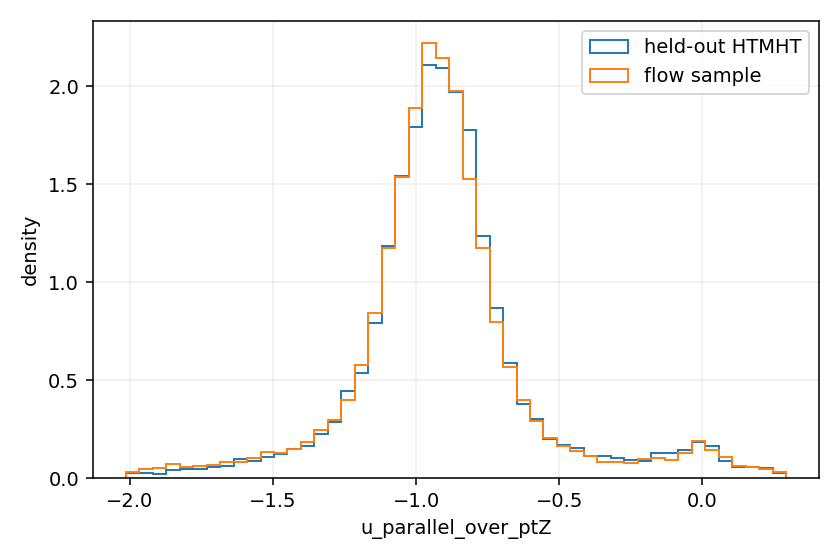}\\[6pt]
  \includegraphics[width=0.48\textwidth]{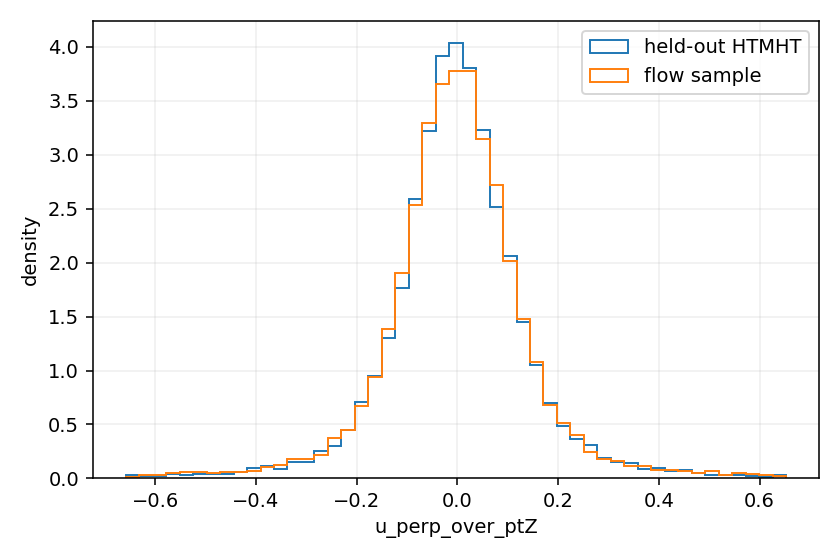}\hfill
  \includegraphics[width=0.48\textwidth]{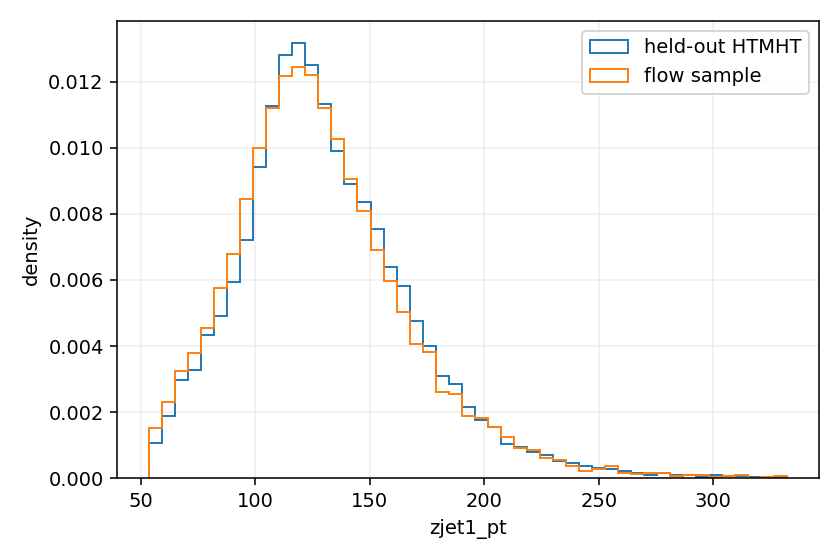}
  \caption{Baseline closure plots for the recoil components and leading $Z$-daughter jet kinematics.}
\end{figure}

\begin{figure}[p]
  \centering
  \includegraphics[width=0.48\textwidth]{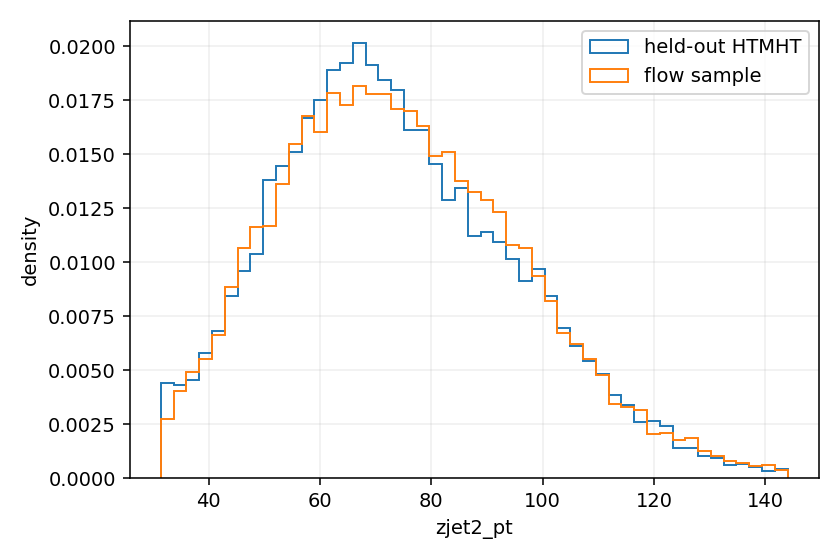}\hfill
  \includegraphics[width=0.48\textwidth]{images/closure_jet1_pt.png}\\[6pt]
  \includegraphics[width=0.48\textwidth]{images/closure_jet2_pt.png}\hfill
  \includegraphics[width=0.48\textwidth]{images/closure_jet2_eta.png}
  \caption{Baseline closure plots for the $Z$-daughter and extra-jet $p_T$ observables.}
\end{figure}

\begin{figure}[p]
  \centering
  \includegraphics[width=0.48\textwidth]{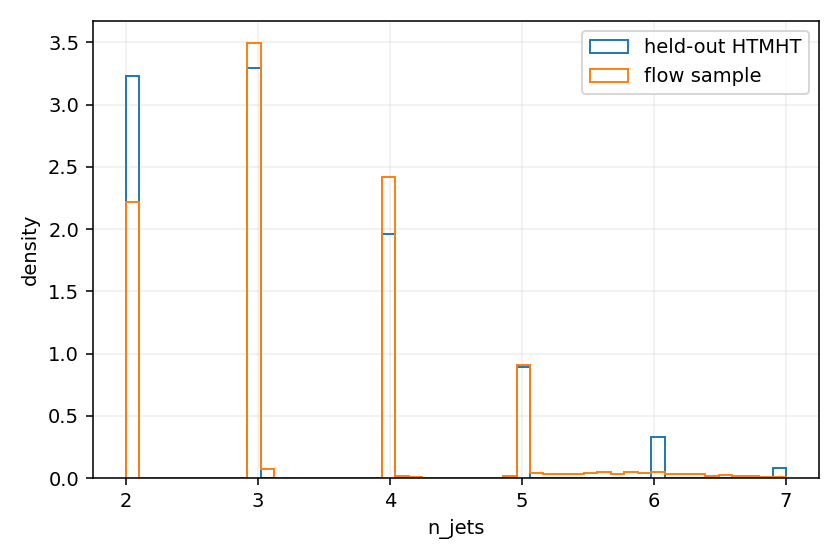}\hfill
  \includegraphics[width=0.48\textwidth]{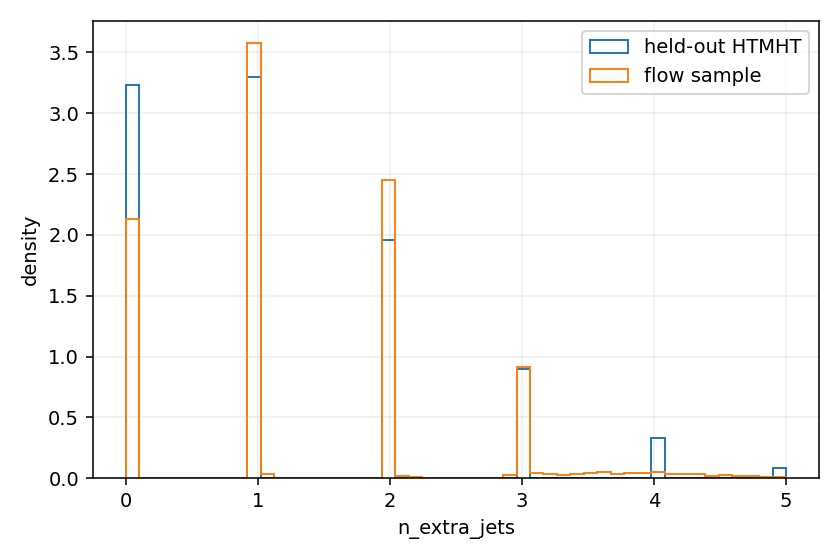}
  \caption{Baseline closure plots for the jet-multiplicity observables.}
\end{figure}

\clearpage
\section{Reproducibility and code availability}
\label{sec:reproducibility}

Full analysis configuration, per-file processing counts, train/validation
split indices, training logs, closure diagnostics, and sensitivity outputs
are available with the accompanying analysis code. The background extraction
pipeline was implemented in Python 3.12 using \texttt{uproot} and
\texttt{awkward} for I/O and array processing and \texttt{numpy} for
numerical operations; density modelling and sensitivity evaluation used
\texttt{PyTorch} and \texttt{torchdiffeq}. Signal samples were produced
with a reproducible \textsc{MadGraph5\_aMC@NLO} (3.x branch),
\textsc{Pythia8}, and \textsc{Delphes} toolchain, with \textsc{ROOT},
\textsc{HepMC3}, and \textsc{FastJet} installed via \texttt{conda-forge};
the \texttt{DMsimp\_s\_spin1} UFO model was used for dark-matter production.
The Delphes CMS card (\texttt{delphes\_card\_CMS.tcl}) does not include a
pileup overlay module, so \texttt{n\_vertices} and \texttt{rho} are zero
placeholders in the signal samples.

Key hyperparameters and environment variables for the flow-matching training
are: random seed \texttt{SEED=20260730}, batch size \texttt{BATCH\_SIZE=2048},
learning rate \texttt{LEARNING\_RATE=2e-4}, weight decay
\texttt{WEIGHT\_DECAY=1e-5}, maximum epochs \texttt{MAX\_EPOCHS=500},
early-stopping patience \texttt{PATIENCE=6}, validation NLL check interval
\texttt{NLL\_EVERY=5}, ODE solver \texttt{rk4} with step size 0.05 and
tolerances $10^{-3}$, divergence mode \texttt{hutchinson}, and density
batch size \texttt{DENSITY\_BATCH\_SIZE=16}. The sentinel imputation value
is \texttt{NAN\_SENTINEL\_VALUE=-999.0}. The ablation mode is activated by
setting \texttt{DROP\_EXTRA\_JET\_FEATURES=1}.

Per-stage signal-side cutflow counters were added to the signal-generation
notebook but the resulting per-stage CSVs have not been reviewed to diagnose
why the \texttt{axial\_mx10\_mv20} benchmark has a lower raw selection
efficiency than the other two points. The offline post-selection yields
quoted in the text (116, 487, and 848 events) reflect the final selected
populations prior to trigger-proxy application and NLL scoring.

\clearpage
\nocite{*}
\bibliographystyle{unsrt}
\bibliography{references}

@misc{rasineni_nsf_monoz,
  author        = {Rasineni, Hitesh and Chebrolu, Bhavishya},
  title         = {{Mono-Z Dark Matter Search with Neural Spline Flows Using CMS Run 2015D Open Data}},
  year          = {2026},
  eprint        = {2607.13771},
  archivePrefix = {arXiv},
  primaryClass  = {cs.LG},
  doi           = {10.48550/arXiv.2607.13771},
  note          = {\href{https://doi.org/10.48550/arXiv.2607.13771}{doi:10.48550/arXiv.2607.13771}}
}

@misc{cms_htmht_2015d,
  author       = {{CMS Collaboration}},
  title        = {{HTMHT primary dataset in MINIAOD format from RunD of 2015 (/HTMHT/Run2015D-16Dec2015-v1/MINIAOD)}},
  year         = {2021},
  publisher    = {CERN Open Data Portal},
  doi          = {10.7483/OPENDATA.CMS.SRUP.MHPK},
  url          = {https://opendata.cern.ch/record/24125},
  note         = {\href{https://doi.org/10.7483/OPENDATA.CMS.SRUP.MHPK}{doi:10.7483/OPENDATA.CMS.SRUP.MHPK}}
}

@misc{cms_data_policy,
  author       = {{CMS Collaboration}},
  title        = {{CMS data preservation, re-use and open access policy}},
  year         = {2020},
  howpublished = {CMS Document 6032-v3},
  url          = {https://cms-docdb.cern.ch/cgi-bin/PublicDocDB/ShowDocument?docid=6032},
  note         = {\href{https://cms-docdb.cern.ch/cgi-bin/PublicDocDB/ShowDocument?docid=6032}{CMS document 6032-v3}}
}

@misc{cms_lumi_guide,
  author       = {{CMS Open Data}},
  title        = {{Luminosity}},
  year         = {2026},
  howpublished = {CMS Open Data Guide},
  url          = {https://cms-opendata-guide.web.cern.ch/analysis/lumi/},
  note         = {\href{https://cms-opendata-guide.web.cern.ch/analysis/lumi/}{CMS Open Data luminosity guide}}
}

@article{abercrombie_dm_forum,
  author        = {Abercrombie, D. and others},
  title         = {{Dark Matter Benchmark Models for Early LHC Run-2 Searches: Report of the ATLAS/CMS Dark Matter Forum}},
  journal       = {Phys. Dark Univ.},
  volume        = {27},
  pages         = {100371},
  year          = {2020},
  eprint        = {1507.00966},
  archivePrefix = {arXiv},
  primaryClass  = {hep-ex},
  doi           = {10.1016/j.dark.2019.100371},
  note          = {\href{https://doi.org/10.1016/j.dark.2019.100371}{doi:10.1016/j.dark.2019.100371}}
}

@article{backovic_schannel,
  author        = {Backovic, Mihailo and Kramer, Michael and Maltoni, Fabio and Martini, Antony and Mawatari, Kentarou and Pellen, Mathieu},
  title         = {{Higher-order QCD predictions for dark matter production at the LHC in simplified models with s-channel mediators}},
  journal       = {Eur. Phys. J. C},
  volume        = {75},
  number        = {10},
  pages         = {482},
  year          = {2015},
  eprint        = {1508.05327},
  archivePrefix = {arXiv},
  primaryClass  = {hep-ph},
  doi           = {10.1140/epjc/s10052-015-3700-6},
  note          = {\href{https://doi.org/10.1140/epjc/s10052-015-3700-6}{doi:10.1140/epjc/s10052-015-3700-6}}
}

@article{neubert_monoz,
  author        = {Neubert, Matthias and Wang, Jian and Zhang, Cen},
  title         = {{Higher-order QCD predictions for dark matter production in mono-Z searches at the LHC}},
  journal       = {JHEP},
  volume        = {02},
  pages         = {082},
  year          = {2016},
  eprint        = {1509.05785},
  archivePrefix = {arXiv},
  primaryClass  = {hep-ph},
  doi           = {10.1007/JHEP02(2016)082},
  note          = {\href{https://doi.org/10.1007/JHEP02(2016)082}{doi:10.1007/JHEP02(2016)082}}
}

@article{alwall_mg5amc,
  author        = {Alwall, J. and Frederix, R. and Frixione, S. and Hirschi, V. and Maltoni, F. and Mattelaer, O. and Shao, H.-S. and Stelzer, T. and Torrielli, P. and Zaro, M.},
  title         = {{The automated computation of tree-level and next-to-leading order differential cross sections, and their matching to parton shower simulations}},
  journal       = {JHEP},
  volume        = {07},
  pages         = {079},
  year          = {2014},
  eprint        = {1405.0301},
  archivePrefix = {arXiv},
  primaryClass  = {hep-ph},
  doi           = {10.1007/JHEP07(2014)079},
  note          = {\href{https://doi.org/10.1007/JHEP07(2014)079}{doi:10.1007/JHEP07(2014)079}}
}

@article{sjostrand_pythia82,
  author        = {Sjostrand, Torbjorn and Ask, Stefan and Christiansen, Jesper R. and Corke, Richard and Desai, Nishita and Ilten, Philip and Mrenna, Stephen and Prestel, Stefan and Rasmussen, Christine O. and Skands, Peter Z.},
  title         = {{An introduction to PYTHIA 8.2}},
  journal       = {Comput. Phys. Commun.},
  volume        = {191},
  pages         = {159--177},
  year          = {2015},
  eprint        = {1410.3012},
  archivePrefix = {arXiv},
  primaryClass  = {hep-ph},
  doi           = {10.1016/j.cpc.2015.01.024},
  note          = {\href{https://doi.org/10.1016/j.cpc.2015.01.024}{doi:10.1016/j.cpc.2015.01.024}}
}

@article{delphes3,
  author        = {de Favereau, J. and Delaere, C. and Demin, P. and Giammanco, A. and Lema{\^i}tre, V. and Mertens, A. and Selvaggi, M.},
  collaboration = {DELPHES 3},
  title         = {{DELPHES 3, A modular framework for fast simulation of a generic collider experiment}},
  journal       = {JHEP},
  volume        = {02},
  pages         = {057},
  year          = {2014},
  eprint        = {1307.6346},
  archivePrefix = {arXiv},
  primaryClass  = {hep-ex},
  doi           = {10.1007/JHEP02(2014)057},
  note          = {\href{https://doi.org/10.1007/JHEP02(2014)057}{doi:10.1007/JHEP02(2014)057}}
}

@article{cacciari_antikt,
  author        = {Cacciari, Matteo and Salam, Gavin P. and Soyez, Gregory},
  title         = {{The anti-k(t) jet clustering algorithm}},
  journal       = {JHEP},
  volume        = {04},
  pages         = {063},
  year          = {2008},
  eprint        = {0802.1189},
  archivePrefix = {arXiv},
  primaryClass  = {hep-ph},
  doi           = {10.1088/1126-6708/2008/04/063},
  note          = {\href{https://doi.org/10.1088/1126-6708/2008/04/063}{doi:10.1088/1126-6708/2008/04/063}}
}

@article{cowan_asymptotic,
  author        = {Cowan, Glen and Cranmer, Kyle and Gross, Eilam and Vitells, Ofer},
  title         = {{Asymptotic formulae for likelihood-based tests of new physics}},
  journal       = {Eur. Phys. J. C},
  volume        = {71},
  pages         = {1554},
  year          = {2011},
  eprint        = {1007.1727},
  archivePrefix = {arXiv},
  primaryClass  = {physics.data-an},
  doi           = {10.1140/epjc/s10052-011-1554-0},
  note          = {\href{https://doi.org/10.1140/epjc/s10052-011-1554-0}{doi:10.1140/epjc/s10052-011-1554-0}}
}

@article{lipman_flow_matching,
  author        = {Lipman, Yaron and Chen, Ricky T. Q. and Ben-Hamu, Heli and Nickel, Maximilian and Le, Matt},
  title         = {{Flow Matching for Generative Modeling}},
  journal       = {International Conference on Learning Representations},
  year          = {2023},
  eprint        = {2210.02747},
  archivePrefix = {arXiv},
  primaryClass  = {cs.LG},
  doi           = {10.48550/arXiv.2210.02747},
  note          = {\href{https://doi.org/10.48550/arXiv.2210.02747}{doi:10.48550/arXiv.2210.02747}}
}

@inproceedings{chen_neural_ode,
  author        = {Chen, Ricky T. Q. and Rubanova, Yulia and Bettencourt, Jesse and Duvenaud, David K.},
  title         = {{Neural Ordinary Differential Equations}},
  booktitle     = {Advances in Neural Information Processing Systems},
  volume        = {31},
  year          = {2018},
  eprint        = {1806.07366},
  archivePrefix = {arXiv},
  primaryClass  = {cs.LG},
  note          = {\href{https://doi.org/10.48550/arXiv.1806.07366}{doi:10.48550/arXiv.1806.07366}}
}

@inproceedings{grathwohl_ffjord,
  author        = {Grathwohl, Will and Chen, Ricky T. Q. and Bettencourt, Jesse and Sutskever, Ilya and Duvenaud, David},
  title         = {{FFJORD: Free-form Continuous Dynamics for Scalable Reversible Generative Models}},
  booktitle     = {International Conference on Learning Representations},
  year          = {2019},
  eprint        = {1810.01367},
  archivePrefix = {arXiv},
  primaryClass  = {cs.LG},
  note          = {\href{https://doi.org/10.48550/arXiv.1810.01367}{doi:10.48550/arXiv.1810.01367}}
}

@inproceedings{durkan_nsf,
  author        = {Durkan, Conor and Bekasov, Artur and Murray, Iain and Papamakarios, George},
  title         = {{Neural Spline Flows}},
  booktitle     = {Advances in Neural Information Processing Systems},
  volume        = {32},
  year          = {2019},
  eprint        = {1906.04032},
  archivePrefix = {arXiv},
  primaryClass  = {stat.ML},
  note          = {\href{https://doi.org/10.48550/arXiv.1906.04032}{doi:10.48550/arXiv.1906.04032}}
}

@article{papamakarios_flows,
  author        = {Papamakarios, George and Nalisnick, Eric and Rezende, Danilo Jimenez and Mohamed, Shakir and Lakshminarayanan, Balaji},
  title         = {{Normalizing Flows for Probabilistic Modeling and Inference}},
  journal       = {J. Mach. Learn. Res.},
  volume        = {22},
  number        = {57},
  pages         = {1--64},
  year          = {2021},
  eprint        = {1912.02762},
  archivePrefix = {arXiv},
  primaryClass  = {stat.ML},
  note          = {\href{https://doi.org/10.48550/arXiv.1912.02762}{doi:10.48550/arXiv.1912.02762}}
}

@misc{lassila_cms_opendata,
  author        = {Lassila-Perini, Kati and others},
  title         = {{Using CMS Open Data in research --- challenges and directions}},
  year          = {2021},
  eprint        = {2106.05726},
  archivePrefix = {arXiv},
  primaryClass  = {hep-ex},
  note          = {\href{https://doi.org/10.48550/arXiv.2106.05726}{doi:10.48550/arXiv.2106.05726}}
}

@article{bellis_cms_dp,
  author        = {Bellis, Matthew and McCauley, Thomas},
  title         = {{CMS Data Preservation and Open Access: status and plans}},
  journal       = {Data Preservation in High Energy Physics: Global Report},
  year          = {2026},
  eprint        = {2607.06775},
  archivePrefix = {arXiv},
  primaryClass  = {hep-ex},
  note          = {\href{https://doi.org/10.48550/arXiv.2607.06775}{doi:10.48550/arXiv.2607.06775}}
}

@article{kraml_spin2,
  author        = {Kraml, Sabine and Laa, Ulrich and Mawatari, Kentarou and Yamashita, Kenta},
  title         = {{Simplified dark matter models with a spin-2 mediator at the LHC}},
  journal       = {Eur. Phys. J. C},
  volume        = {77},
  number        = {5},
  pages         = {326},
  year          = {2017},
  eprint        = {1701.07008},
  archivePrefix = {arXiv},
  primaryClass  = {hep-ph},
  doi           = {10.1140/epjc/s10052-017-4871-0},
  note          = {\href{https://doi.org/10.1140/epjc/s10052-017-4871-0}{doi:10.1140/epjc/s10052-017-4871-0}}
}

@misc{lhc_dm_wg_heavy,
  author       = {{LHC Dark Matter Working Group}},
  title        = {{Comparing LHC searches for heavy mediators of dark matter production in visible and invisible decay channels}},
  year         = {2017},
  eprint       = {1703.05703},
  archivePrefix = {arXiv},
  primaryClass = {hep-ex},
  doi          = {10.48550/arXiv.1703.05703},
  note         = {\href{https://doi.org/10.48550/arXiv.1703.05703}{doi:10.48550/arXiv.1703.05703}}
}

@article{bell_busoni_scalar,
  author        = {Bell, Nicole F. and Busoni, Giorgio and Sanderson, Isaac W.},
  title         = {{Self-consistent Dark Matter Simplified Models with an s-channel scalar mediator}},
  journal       = {JCAP},
  volume        = {03},
  pages         = {015},
  year          = {2017},
  eprint        = {1612.03475},
  archivePrefix = {arXiv},
  primaryClass  = {hep-ph},
  doi           = {10.1088/1475-7516/2017/03/015},
  note          = {\href{https://doi.org/10.1088/1475-7516/2017/03/015}{doi:10.1088/1475-7516/2017/03/015}}
}

@article{cacciari_fastjet,
  author        = {Cacciari, Matteo and Salam, Gavin P. and Soyez, Gregory},
  title         = {{FastJet User Manual}},
  journal       = {Eur. Phys. J. C},
  volume        = {72},
  pages         = {1896},
  year          = {2012},
  eprint        = {1111.6097},
  archivePrefix = {arXiv},
  primaryClass  = {hep-ph},
  doi           = {10.1140/epjc/s10052-012-1896-2},
  note          = {\href{https://doi.org/10.1140/epjc/s10052-012-1896-2}{doi:10.1140/epjc/s10052-012-1896-2}}
}

@article{nnpdf3,
  author        = {Ball, Richard D. and others},
  collaboration = {NNPDF},
  title         = {{Parton distributions for the LHC Run II}},
  journal       = {JHEP},
  volume        = {04},
  pages         = {040},
  year          = {2015},
  eprint        = {1410.8849},
  archivePrefix = {arXiv},
  primaryClass  = {hep-ph},
  doi           = {10.1007/JHEP04(2015)040},
  note          = {\href{https://doi.org/10.1007/JHEP04(2015)040}{doi:10.1007/JHEP04(2015)040}}
}

@article{planck2018_vi,
  author        = {Aghanim, N. and others},
  collaboration = {Planck},
  title         = {{Planck 2018 results. VI. Cosmological parameters}},
  journal       = {Astron. Astrophys.},
  volume        = {641},
  pages         = {A6},
  year          = {2020},
  eprint        = {1807.06209},
  archivePrefix = {arXiv},
  primaryClass  = {astro-ph.CO},
  doi           = {10.1051/0004-6361/201833910},
  note          = {\href{https://doi.org/10.1051/0004-6361/201833910}{doi:10.1051/0004-6361/201833910}}
}

@inproceedings{dinh_realnvp,
  author    = {Dinh, Laurent and Sohl-Dickstein, Jascha and Bengio, Samy},
  title     = {{Density estimation using Real NVP}},
  booktitle = {International Conference on Learning Representations},
  year      = {2017},
  eprint    = {1605.08803},
  archivePrefix = {arXiv},
  primaryClass  = {cs.LG},
  note      = {\href{https://doi.org/10.48550/arXiv.1605.08803}{doi:10.48550/arXiv.1605.08803}}
}

@inproceedings{kingma_glow,
  author    = {Kingma, Durk P. and Dhariwal, Prafulla},
  title     = {{Glow: Generative Flow with Invertible 1x1 Convolutions}},
  booktitle = {Advances in Neural Information Processing Systems},
  volume    = {31},
  year      = {2018},
  eprint    = {1807.03039},
  archivePrefix = {arXiv},
  primaryClass  = {stat.ML},
  note      = {\href{https://doi.org/10.48550/arXiv.1807.03039}{doi:10.48550/arXiv.1807.03039}}
}

@article{nachman_shih_anode,
  author        = {Nachman, Benjamin and Shih, David},
  title         = {{Anomaly Detection with Density Estimation}},
  journal       = {Phys. Rev. D},
  volume        = {101},
  pages         = {075042},
  year          = {2020},
  eprint        = {2001.04990},
  archivePrefix = {arXiv},
  primaryClass  = {hep-ph},
  doi           = {10.1103/PhysRevD.101.075042},
  note          = {\href{https://doi.org/10.1103/PhysRevD.101.075042}{doi:10.1103/PhysRevD.101.075042}}
}

@article{hallin_cathode,
  author        = {Hallin, Anna and Kasieczka, Gregor and Kraml, Sabine and Lessa, Andre and Shih, David},
  title         = {{Classifying anomalies through outer density estimation}},
  journal       = {Phys. Rev. D},
  volume        = {106},
  pages         = {055006},
  year          = {2022},
  eprint        = {2109.00546},
  archivePrefix = {arXiv},
  primaryClass  = {hep-ph},
  doi           = {10.1103/PhysRevD.106.055006},
  note          = {\href{https://doi.org/10.1103/PhysRevD.106.055006}{doi:10.1103/PhysRevD.106.055006}}
}

@article{metodiev_cwola,
  author        = {Metodiev, Eric M. and Nachman, Benjamin and Thaler, Jesse},
  title         = {{Classification without labels: Learning from mixed samples in high energy physics}},
  journal       = {JHEP},
  volume        = {10},
  pages         = {174},
  year          = {2017},
  eprint        = {1708.02949},
  archivePrefix = {arXiv},
  primaryClass  = {hep-ph},
  doi           = {10.1007/JHEP10(2017)174},
  note          = {\href{https://doi.org/10.1007/JHEP10(2017)174}{doi:10.1007/JHEP10(2017)174}}
}

\end{document}